\documentclass[twocolumn,showpacs,amsmath,amssymb,superscriptaddress,floatfix]{revtex4-2}
\usepackage{graphicx,url,hyperref,dcolumn,bm,lineno}
\hypersetup{colorlinks,citecolor=blue,linkcolor=red,urlcolor=blue}

\usepackage{comment}
\usepackage[dvipsnames]{xcolor}
\usepackage{wasysym}
\usepackage[normalem]{ulem}
\usepackage{soul}

\modulolinenumbers[5]
\begin{document}
%--------------------------------------------------------------------
\title{Coordination-number dependent universality in Mixed-Wet Percolation}
%--------------------------------------------------------------------

\author{Jnana Ranjan Das}
\email{d.jnana@iitg.ac.in}
\affiliation{Department of Physics, Indian Institute of Technology Guwahati, Guwahati-781039 Assam, India}

\author{Santanu Sinha}
\email{santanu.sinha@ntnu.no}
\affiliation{PoreLab, Department of Physics, Norwegian University of Science and Technology, NO--7491 Trondheim, Norway}

\author{Alex Hansen}
\email{alex.hansen@ntnu.no}
\affiliation{PoreLab, Department of Physics, Norwegian University of Science and Technology, NO--7491 Trondheim, Norway}

\author{Sitangshu B.\ Santra}
\email{santra@iitg.ac.in}
\affiliation{Department of Physics, Indian Institute of Technology Guwahati, Guwahati-781039 Assam, India}

\date{\today {}}

%--------------------------------------------------------------------
\date{\today {}}
%--------------------------------------------------------------------
\begin{abstract}
Mixed-wet percolation was introduced recently in the context of two-phase flow in porous media. In this model, the sites of the primal lattice are occupied with a certain probability $p$, and bonds are placed on the dual lattice between two adjacent occupied and unoccupied sites of the primal lattice. The occupied bonds on the dual lattice form perimeter bond clusters. We investigate here the scaling properties of the geometric quantities associated with the perimeter bond clusters of mixed-wet percolation on the dual triangular and dual honeycomb lattices in two dimensions. The perimeter bond clusters on the triangular lattice represent the boundaries of the site clusters on the primal lattice, whereas the perimeter bond clusters on the honeycomb lattice represent their hulls. Because of the low coordination number $z$ of the honeycomb lattice, the external and internal perimeter bond clusters remain isolated. Although mixed-wet percolation on the dual triangular lattice with a higher coordination number ($z=6$) exhibits ordinary cluster percolation, the model on the dual honeycomb lattice with a lower coordination number ($z=3$) exhibits hull percolation. Such a $z$ dependent crossover of universality in mixed-wet percolation is rare in the percolation literature. However, the external and internal perimeter bond clusters associated with a site cluster combine to form cluster envelopes. These cluster envelopes belong to the cluster percolation universality class.
\end{abstract}

%--------------------------------------------------------------------
\maketitle
%--------------------------------------------------------------------
%\linenumbers
%--------------------------------------------------------------------
\section{Introduction}
\label{intro}
The universality class of a statistical system depends on the dimensionality of space $d$ and the number of components in the order parameter \cite{stanley1987introduction,yeomans1992statistical,binney1992theory,goldenfeld2018lectures}. In a percolation model, the order parameter, the probability that a lattice site belongs to the spanning cluster, is a scalar quantity \cite{broadbent1957percolation,stauffer2018introduction,christensen2005complexity,bunde2012fractals,sorensen,bottcher2021computational}. Hence, most of the percolation models studied in a particular space dimension belong to the percolation universality class \cite{stauffer2018introduction,bunde2012fractals} in the same spatial dimension. We investigate a two-dimensional percolation model, namely Mixed-Wet percolation, that exhibits a crossover in universality depending on coordination number $z$, which is rare in percolation literature.

%the bond-cluster properties of a recently introduced
%We will demonstrate that such a breakdown depends on the lattice coordination number.
%The probability of occupation of the site (or bond) $p$ is the only parameter in a percolation problem \cite{broadbent1957percolation,stauffer2018introduction,christensen2005complexity,bunde2012fractals,sorensen,bottcher2021computational}. So, the cluster properties of the percolation model studied on different lattices in $d=2$ dimensions belong to the universality class of cluster percolation \cite{stauffer2018introduction,bunde2012fractals}. In this paper, we present a breakdown of universality on a {\color{red} recently introduced model} called Mixed-Wet Percolation (MWP). 

Mixed-Wet Percolation has its name from immiscible two-phase flow in porous media \cite{heiba1983effect,sahimi2011flow,blunt2017multiphase,feder2022physics,ltz88,lz89,ga19}, where the porous medium consists of a mixture of two types of grains with two different wetting properties \cite{irannezhad2023fluid,irannezhad2023characteristics}. Imagine a Hele-Shaw cell filled with a single-layer random mixture of these grains. The space between the grains forms a porous network in which two fluids and their interfaces will move. Each link in this network will be adjacent to two grains. If fluid $A$ is wetting with respect to grain type $1$ and fluid $B$ is wetting with respect to grain type $2$, there are three situations: $(i)$ If both grains are of type $1$ and an interface between the two fluids passes through the link, the capillary force generated by the curvature of the interface will point towards fluid $B$, which is non-wetting in this case. $(ii)$ If the two grains are of type $2$, the capillary force will point towards fluid $A$, which is the non-wetting one in this case. $(iii)$ If the grains are of different types, the curvature of the interface will be much smaller, and we may ideally assume it to disappear: there is no capillary force associated with the interface in this case \cite{fyhn2021rheology,fyhn2023effective}. The percolation model we study here considers the cluster structure of pore throats that do not exhibit capillary forces when fluid interfaces are present. However, in the flow problem, only a fraction of the pore throats will at any time contain interfaces. This strongly influences the flow problem and needs to be considered \cite{fyhn2023effective}. Thus, the percolation problem we study must be combined with the motion of the fluids for a full description of the experimental system.         

The percolation properties of the two-phase flow are relevant at low capillary numbers, where capillary forces dominate over viscous forces, and the flow is governed by disorder in the properties of the pore-space, such as pore sizes and wettability \cite{ltz88,lz89,sahimi2011flow}.  Transforming the mixed-wet two-phase flow problem into a formal percolation problem, we imagine the grains being placed at the nodes of a given lattice. We refer to this as the primal lattice (PL). The porous network between the grains may then be modelled by the dual lattice (DL) associated with the primal lattice. The two types of grains, type $1$ and $2$, may then be modelled as the nodes of the primal lattice being either occupied or empty. A bond on the dual lattice is then only placed on a link if the adjacent nodes are of different types, one occupied and one empty. The clusters formed by these bonds are free of capillary forces.   
 
Recently, MWP was introduced on the square lattice, the self-dual lattice, with a coordination number $z_s=4$, where we found a new cluster geometry at a much lower percolation threshold than that of ordinary percolation, whose critical scaling properties belong to the ordinary cluster percolation universality in two dimensions \cite{DAS2025130957}. Here, we consider triangular and honeycomb lattices with coordination numbers $z_t=6$ and $z_h=3$, respectively, in two dimensions. To study MWP on these two lattices, once we have used a honeycomb lattice as the primal lattice, in which bond clusters form on the dual triangular lattice ($z_t=6$), and then we have used a triangular lattice as the primal lattice, with bond clusters forming on its dual honeycomb lattice ($z_h=3$). Interestingly, not only are the outcomes markedly different between these two apparently equivalent processes on lattices with two distinct $z$ values, but also their critical properties belong to two different universality classes.

%Thus, the primal and dual lattices have the same coordination number $z_{\rm s}=4$. The site clusters in the primal lattice have one-to-one correspondence with the bond clusters in the dual lattice. It is found that the universality class of the bond clusters belongs to ordinary percolation \cite{DAS2025130957}. 
%\st{ In the first case, the bond clusters form on the triangular (dual) lattice, whereas in the latter case, they form on the honeycomb (dual) lattice, and they have very different coordination numbers.}
We present MWP on both pairs of primal-dual lattices (honeycomb-triangular, triangular-honeycomb) in Section \ref{model}. The results are presented in Section \ref{Results}.  
%We start by considering the wrapping probability in Sub-section \ref{Wrapping}, which identifies spanning clusters in biperiodic systems.  We then go on to study the cluster size distribution in Sub-section \ref{cluster_size_dis}, the order parameter in Sub-section \ref{FSS} and fluctuations of the order parameter in Sub-section \ref{FOP}. Sub-section \ref{fractal} discusses fractal dimensions, and we end the Results Section by investigating the external and internal perimeters of the clusters, see Sub-section \ref{dhl_ext}. 
We summarize and discuss our findings in Section \ref{conclusion}.
%--------------------------------------------------------------------
\section{Model}
\label{model}

\begin{figure*}[ht]
    \centering{\hfill
        \includegraphics[width=0.34\linewidth,clip]{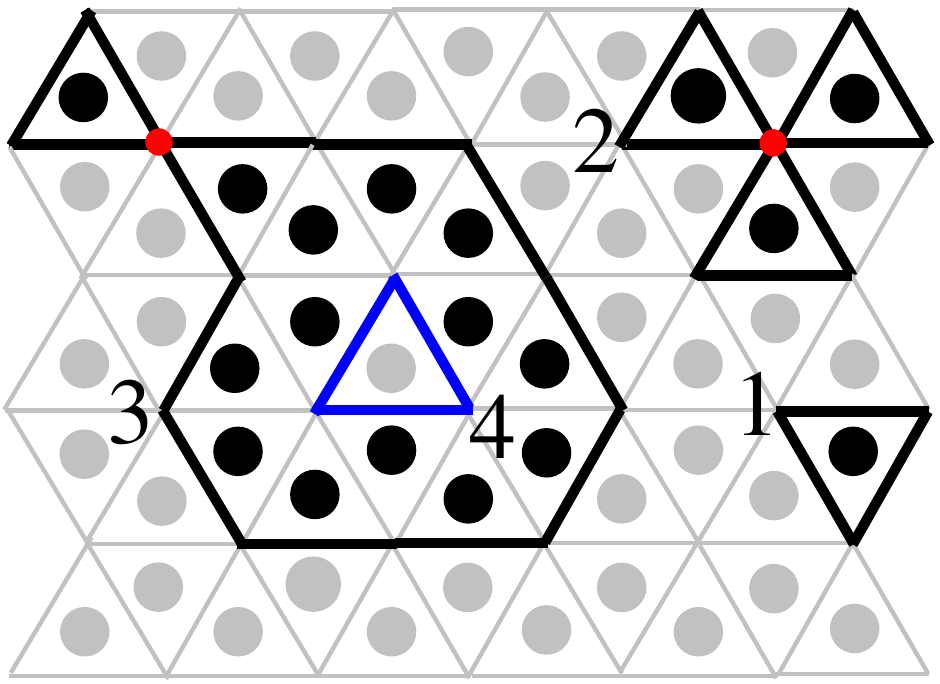}\hfill
        \includegraphics[width=0.3\linewidth,clip]{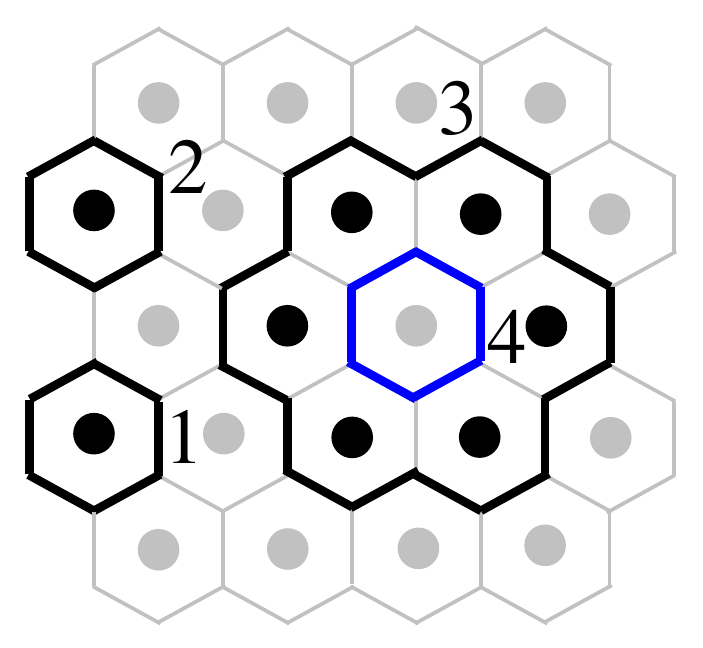}\hfill
         \includegraphics[width=0.35\linewidth]{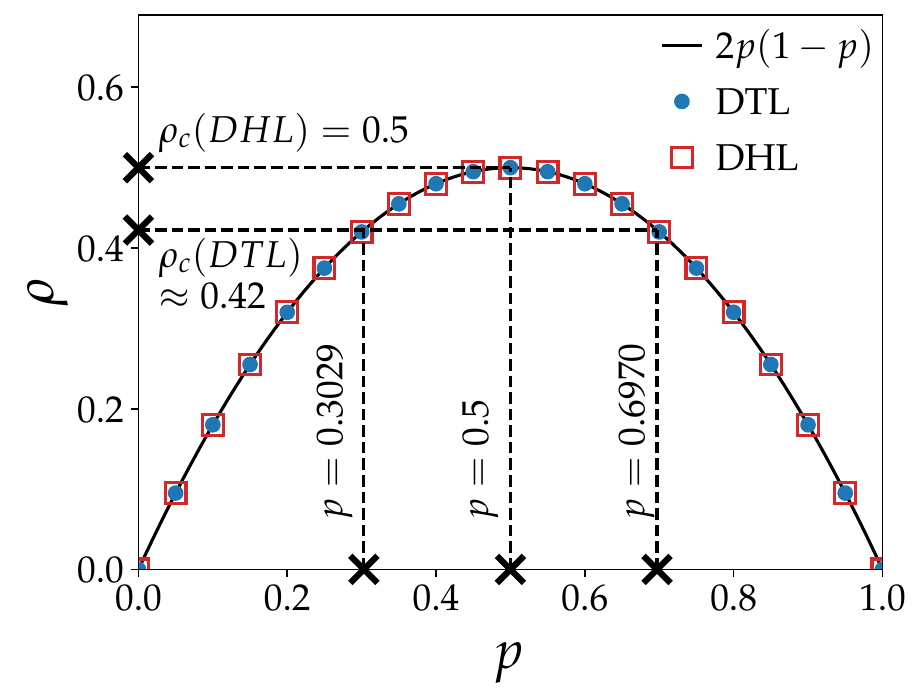}\hfill}\\
        \centering
        (a) DTL \hspace{0.25\linewidth} (b) DHL \hspace{0.25\linewidth} (c) bond density \\
       \caption{\label{clusters} Typical illustrations of MWP on the (a) DTL and (b) DHL. The black circles represent occupied sites, whereas the grey circles represent unoccupied sites on the PL. The thin grey lines indicate the links on the DL, and the thick lines (shown in black or blue) indicate occupied bonds on the DL. The detached inner perimeter clusters are shown in blue. The red bullets in (a) indicate knots on the DL. (c) Plot of the bond density $\rho$ as a function of $p$. The blue bullets and red squares represent the bond density measured from the simulation of the DTL and DHL models of lattice size $L = 512$, respectively. The solid black line corresponds to the relation $\rho=2p(1-p)$. The dashed lines with crosses on the $p$-axis at $p=0.3029$ and $1-p=0.6970$ indicate the two thresholds of the DTL, while $p = 0.5$ indicates the threshold of the DHL. Critical bond densities $\rho_c$ corresponding to these $p_c$'s are marked on the $\rho$-axis.}
\end{figure*}

Here, we consider a pair of triangular and honeycomb lattices, as they are dual to each other, to study MWP. First, we take the honeycomb lattice as the primal lattice (PL), whose dual lattice (DL) is the triangular lattice, which we refer to as the dual-triangular lattice (DTL). Secondly, we take the triangular lattice as the PL, whose DL is the honeycomb lattice, which we refer to as the dual-honeycomb lattice (DHL). Each site on the PL is independently occupied with a certain probability $p$ (or remains unoccupied with probability $1-p$). These occupied and unoccupied sites represent the two types of grains in the porous media. A bond is placed with unit probability on the links of the DL that appear between any adjacent pair of randomly occupied and unoccupied sites on the PL. A few finite clusters that appear in DTL and DHL models are shown in Fig. \ref{clusters}(a)  and \ref{clusters}(b), respectively. The black circles represent the randomly occupied sites, and the grey circles represent the remaining unoccupied sites on the PL. The thin grey lines represent the links on the DL. The thick black (or blue) lines represent the bonds occupying the links with unit probability on the DL (between the occupied and unoccupied sites on the PL). There are two possible perimeter bond clusters: one is a boundary of occupied sites (in black), and the other is a boundary of unoccupied sites (in blue). If two perimeter clusters of the same type, with sizes $b_1$ and $b_2$, meet at a DL site, they form a single perimeter bond cluster of size $b=b_1+b_2$. We call such a site a knot shown by a red bullet. Four perimeter bond clusters on the DTL are shown in Fig. \ref{clusters}(a). The cluster $1$ is an elementary three-bond perimeter cluster on the DTL around a single occupied site on the PL. Cluster $2$ is a combination of three elementary perimeters on the DTL connected by a knot. It is then considered a single perimeter bond cluster of nine bonds around three occupied sites on the PL, each separated by the next-nearest-neighbour distance. Cluster $3$ is a combination of a perimeter of size $b_1=10$ and an elementary perimeter of size $b_2=3$ by a knot on the DTL. It is now a cluster of size $b=13$. In this case, the two occupied sites on either side of the knot are separated by the next-to-next-nearest neighbour distance on the PL. It should be noted that merging of two perimeters by a knot can couple two occupied (or unoccupied) sites on the PL separated by at most next-to-next-nearest neighbour distance, leading to next-to-next-nearest neighbour site percolation on the PL. Cluster $4$ is a perimeter enclosing an unoccupied site (or a hole) within a cluster of occupied sites on the PL. It can also be considered the inner perimeter of the cluster of occupied sites, completely detached from the outer perimeter. Though such completely detached inner perimeters occur only rarely on the DTL, if they appear, they will be considered as independent perimeter bond clusters. Perimeters with more complex structure appear involving knots that connect next-nearest neighbour distance and next-to-next-nearest neighbour distance sites on the PL. Every perimeter bond cluster on the DTL then encloses a corresponding site cluster on the PL with sites connected by at most next-to-next-nearest neighbour distance. As a bond cluster percolates on the DTL, the corresponding site cluster, with connectivity up to next-to-next-nearest neighbour distance, also percolates.

%-------------------------------------------------------------------
%--------------------------------------------------------------------

Four perimeter clusters on the DHL are shown in Fig. \ref{clusters}(b). Clusters $1$ and $2$ are elementary perimeters around occupied sites, each of size six. Cluster $3$ is a perimeter of size eighteen around occupied sites. Again, the cluster $4$ is a perimeter around an unoccupied site (or hole) inside a cluster of occupied sites. It is also the inner perimeter of the cluster of occupied sites. In the case of DHL, however, the perimeter bond clusters are simple perimeters without knots, but we continue to refer to them as perimeter bond clusters. They appear around the site clusters with nearest neighbour connectivity only in triangular PL, as shown in Fig. \ref{clusters}(b). Since the two perimeter bond clusters on the DHL (in Fig. \ref{clusters}(b)) never meet, they remain separated at least by the nearest neighbour distance. To have a knot, at least two perimeter bond clusters must meet at a DL site, or at least four bonds must pass through a knot. Since the honeycomb lattice has a coordination number $z_h=3$, it is impossible for four bonds to meet at any lattice site, and thus knots do not arise on the DHL. Without these knots, the perimeter bond clusters cannot have fjords and deep inlets as they would appear with a perimeter bond cluster in DTL (or on the square lattice). Hence, all perimeter bond clusters, either around occupied sites or unoccupied sites, will remain isolated from each other in DHL and will be considered as independent perimeter bond clusters. Thus, the MWP in DTL and DHL is expected to exhibit different percolation properties. 
%{\color{red} Some work closely related to the perimeter clusters on the DHL has been reported previously \cite{Saleur1987,weinrib1985new}. In these studies, site percolation clusters  were generated on a triangular lattice, and their hulls were investigated. In particular, the fractal dimension of the hulls was determined. }

At a given $p$, the average bond density $\rho$, on DTL or DHL, can be estimated as 
\begin{equation}
    \label{rho}
    \rho=\frac{\langle B(p)\rangle}{3L^2}
\end{equation}
where $\langle B(p)\rangle$ is the ensemble-averaged number of occupied bonds on the DL corresponding to the given $p$, and $L$ is the linear size of the triangular lattice. For the triangular and the honeycomb lattices, the total number of links is given by \(z_t L^2/d = 2z_h L^2/d= 3L^2\) as $z_t = 2z_h$ and $d=2$ is the space dimension. Note that on the square lattice, the denominator of Eq. \ref{rho} is $2L^2$ \cite{DAS2025130957}. However, on the DL, a bond always appears between a pair of adjacent occupied and unoccupied sites on the PL. Consequently, the bond density $\rho$ is related to the site occupation probability $p$ on the PL as, 
\begin{equation}
    \label{rho_p}
    \rho=2p(1-p),
\end{equation}
where the factor $2$ appears because a bond can be placed in two ways on DL by interchanging the positions of occupied and unoccupied sites. Note that, the value of bond density $\rho$ for a given $p$ is independent of the lattice structure. In Fig. \ref{clusters}(c), the bond density $\rho$ on the DL of both models is plotted against the site occupation probability $p$ on the PL. The blue bullets and red squares denote the measured values of $\rho$, obtained from Eq. \ref{rho}, for the corresponding values of $p$ for the DTL and DHL, respectively. The solid black line corresponds to the relation in Eq. \ref{rho_p}. It can be seen that the values of $\rho$ obtained from the numerical estimate (Eq. \ref{rho}) and those from the analytic estimate (Eq. \ref{rho_p}) are in agreement. The maximum bond density $\rho =1/2$ is found to be at $p = 1/2$. This means the bond density in MWP never reaches $1$ even if $p=1$, unlike ordinary bond percolation. It is also symmetric about $p = 1/2$. For $p \ll 1/2$, the perimeter bond clusters appear mostly around the occupied site clusters (see Fig. \ref{clusters}(a), (b)), and similarly, for $p \gg 1/2$, they will be mostly around the unoccupied site clusters on the PL. Furthermore, if there is a percolation threshold $p_c<1/2$ at which a perimeter bond cluster spans the dual lattice, then there must be another threshold at $1-p_c>1/2$. If the threshold occurs at $p_c = 1/2$, then the model has only one threshold.

%----------------------------------------------------------
\begin{figure*}[ht]
  \centering{\includegraphics[width=0.48\linewidth]{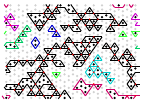} \hfill
    \includegraphics[width=0.48\linewidth]{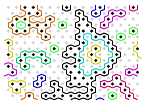}}\\
    \centering
	(a) DTL \hspace{0.48\linewidth} (b) DHL\\     
 \caption{\label{morphology} A typical morphology of clusters on (a) DTL with $p = 0.28$ and (b) DHL with $p = 0.35$ on a lattice of size $L = 16$. The black and grey circles indicate occupied and unoccupied sites on the PL, respectively. The red bullets on the DL in (a) denote the knots. The colours are assigned to the perimeter bond cluster based on their size. The largest cluster is highlighted in black.}
\end{figure*}
%----------------------------------------------------------

\section{Results and Discussion}
\label{Results}
We perform an extensive computer simulation to study MWP on both pairs of lattices by varying the site occupation probability $p$ on the PL, for lattice sizes ranging from $L = 200$ to $L = 2000$ in steps of $200$. Periodic boundary conditions are applied in both horizontal and vertical directions. Perimeter bond clusters are generated on the DL. To determine the size $b$ of a perimeter bond cluster, the burning algorithm \cite{bunde2012fractals} is employed. Here we search the DL site by site, starting from a corner, to identify whether any unburnt bond is associated with it. If an unburnt bond is detected, we burn it and its index is added to a list. At the same time, a counter to record the size $b$ of the bond cluster is incremented by one. Sitting on this burnt bond, we search for unburnt nearest-neighbour bonds at both ends of this root bond. If any unburnt bond is found, it is burned and added to the list. Whenever a bond is burnt, the counter is incremented by one. Each bond in the list is visited sequentially. This burning process continues as long as the list is not exhausted. The entire perimeter bond cluster is identified at the end of the burning process. The value of the counter is the size $b$ of the perimeter bond cluster. Once the burning of a cluster is complete, the search for the next unburnt perimeter bond cluster resumes. The algorithm terminates when no unburnt perimeter bond cluster remains in the entire lattice. The distribution of perimeter bond cluster sizes is then determined. The morphologies of the perimeter bond clusters generated on DTL and DHL for a lattice of size $L = 16$ are shown in Fig. \ref{morphology}(a) and \ref{morphology}(b), respectively. Applying the above burning algorithm to the cluster configuration in Fig. \ref{morphology}, we obtain nine perimeter bond clusters in DTL and eighteen perimeter bond clusters in DHL. It can be seen that the perimeter bond clusters on the DTL (Fig. \ref{morphology}(a)) are mostly composed of perimeters connected to one another by knots. In contrast, the perimeter bond clusters on the DHL (Fig. \ref{morphology}(b)) are isolated perimeters that do not connect one another.

Statistical properties of the perimeter bond clusters on the DL are estimated by taking the ensemble average over $N = 10^5$ samples for each $p$ along with their statistical errors. Though these properties are functions of $\rho$, the scaling behaviour of them are obtained in terms of $\Delta p=|p-p_c|$ instead of $\Delta\rho =|\rho-\rho_c|$. Since $\rho$ is a function of $p$, the leading order singularity remains unchanged in terms of $\Delta p$ as described in \cite{DAS2025130957}. 

%Note that, by employing the Ziff algorithm \cite{ziff1984generation}, one obtains the hull of the perimeter bond cluster, but not the full perimeter bond cluster. 

%{\color{red} Since the perimeter bond clusters are very different from hulls of percolation clusters in general, we have not used Ziff algorithm \cite{ziff1984generation} in determining the perimeter bond clusters.}

\subsection{Wrapping probability}
\label{Wrapping}

A sample is considered to be spanning if at least one perimeter bond cluster extends across the lattice either in the vertical (top-to-bottom) or in the horizontal (left-to-right) directions \cite{newman2000efficient}. The wrapping probability is the probability that a randomly generated sample out of $N$ independent realisations spans the lattice, given by
\begin{equation}
   W(p, L) = \frac{N_{\rm sp}(p,L)}{N}
\end{equation}
where $N_{\rm sp}(p, L)$ is the number of spanning samples. We study $W(p,L)$ for the perimeter bond clusters on the dual lattice as a function of the site occupation probability $p$ on the primal lattice. The finite-size scaling form of $W(p,L)$ is given by,
\begin{equation}
 W(p,L) = \widetilde{W}\left[(p-p_c)L^{1/\nu}\right]
\label{eq-3-1}
\end{equation}
where $\widetilde{W}[x]$ is the scaling function with the scaled variable $x=(p-p_c)L^{1/\nu}$. The exponent $\nu$ is the critical exponent associated with the correlation length $\xi$ of the system. The spanning of a realisation is verified using the burning algorithm by setting fire to the top (or left) row and checking whether the fire reaches the bottom (or right) row \cite{bottcher2021computational,herrmann1984backbone}. The wrapping probabilities $W(p, L)$ for different $L$ are estimated as a function of $p$, taking data in steps of $\Delta p = 0.0001$ for both DTL and DHL.
%----------------------------------------------------------
\begin{figure*}[ht]
\centerline{\hfill\includegraphics[width=0.33\linewidth,clip]{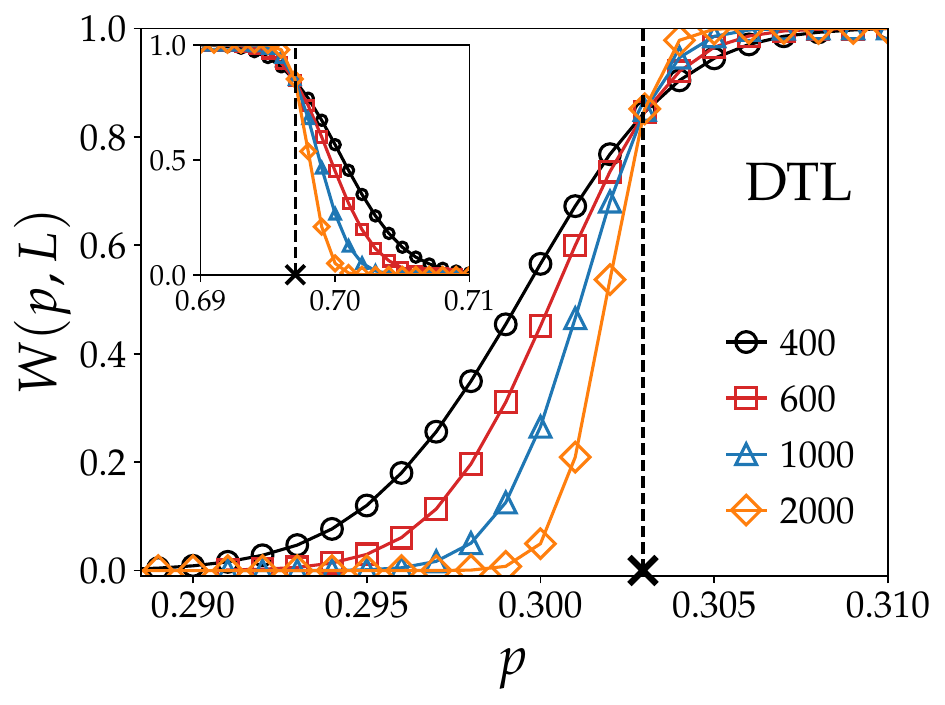}\hfill
            \includegraphics[width=0.33\linewidth]{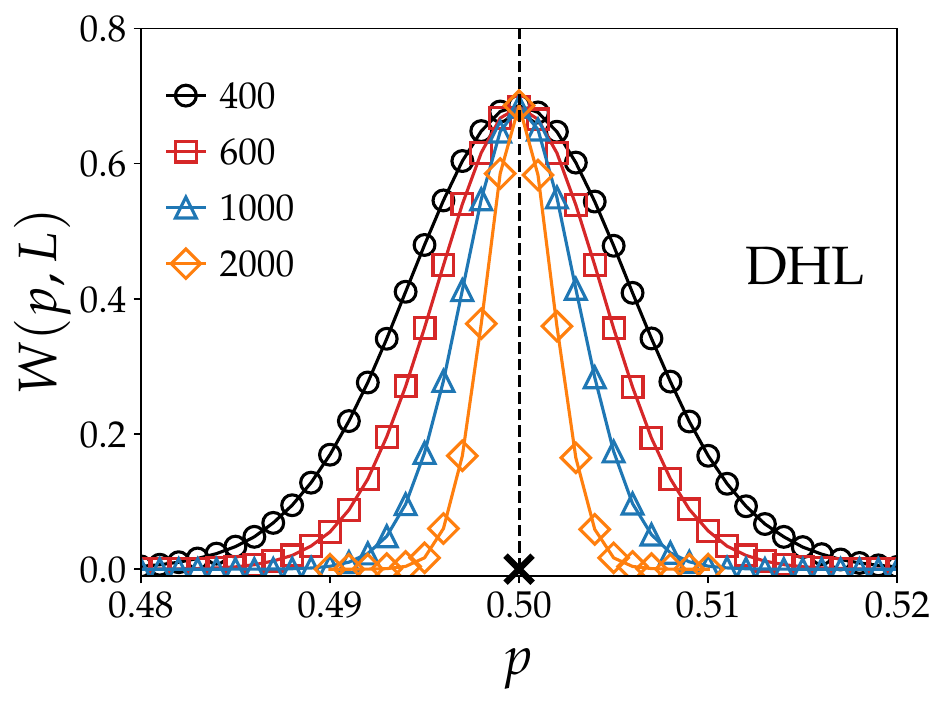}\hfill
            \includegraphics[width=0.33\linewidth]{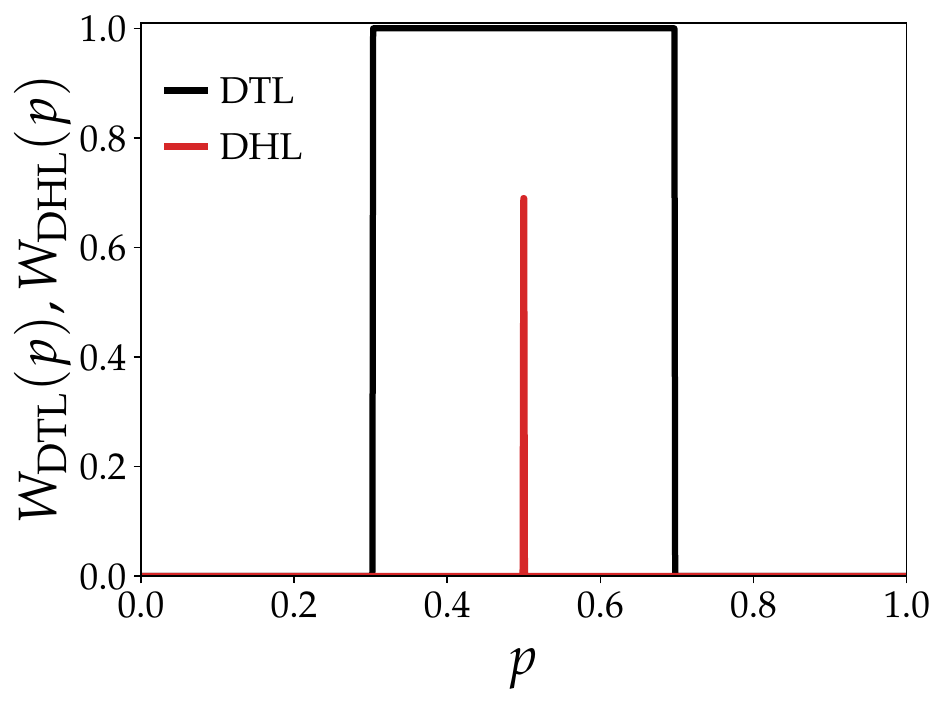}\hfill}
            \centerline{\hfill (a)\hfill\hfill (b)\hfill\hfill (c)\hfill}        
\caption{\label{wp} Plot of $W(p, L)$ versus the occupation probability $p$ is shown for (a) DTL and (b) DHL for various lattice sizes $L$. The dashed line with a cross marked on the $p$-axis indicates the intersection of all the curves at $p = p_c$. (c) Plot of $W_{\rm{DTL}}(p)$ and $W_{\rm{DHL}}(p)$ with lattice size $L \to \infty$ for $p$ varying from $0$ to $1$.}
\end{figure*} 
%---------------------------------------------------------

Now we present the wrapping probability for DTL. In Fig. \ref{wp}(a), $W(p, L)$ is plotted against $p$  for various lattice sizes on DTL. In this plot, we show points in steps of $\Delta p = 0.001$ for clarity. At $p = p_c$, $\widetilde{W}[0]$ is a constant and independent of $L$. Therefore, all $W(p, L)$ curves are expected to intersect each other at $p = p_c$. The curves for different $L$ intersect at a single point, and the corresponding $p$ value is the critical threshold $p_c$. Employing the least square intersection method, the value of $p_c$ is determined as $p_c= 0.3029\pm 0.0001$. The critical threshold $p_c$ is indicated by the black dashed line and marked by a cross on the $p$-axis in Fig. \ref{wp}(a). As the critical threshold $p_c < 1/2$, there exists another critical threshold at $1-p_c = 0.6971$. It is shown in the inset of Fig. \ref{wp}(a) that such a threshold exists. The situation is very similar to the case of the square lattice \cite{DAS2025130957}. Note that $p_c=0.3029$ is also the critical threshold for a site percolation with next-to-next-nearest neighbour distance \cite{malarz2022random,xun2022site}, and $1-p_c = 0.6971$ also corresponds to the critical threshold for a nearest-neighbour site percolation on the honeycomb lattice \cite{suding1999site}. Thus, the critical thresholds for nearest-neighbour $p_c^{\rm NN}$ and next-to-next-nearest neighbour $p_c^{\rm NNNN}$ site percolation on the honeycomb lattice are related through $p_c^{\rm NN} = 1 - p_c^{\rm NNNN}$. To the best of our knowledge, we have not found an explicit report of this relation in the literature. Both the thresholds on the primal lattice correspond to the same critical bond density $\rho_c = 0.4223$ of MWP on DTL, which is far above the ordinary bond percolation critical threshold $p_c = 2 \sin(\pi/18) = 0.34729635533$ \cite{sykes1964exact} on the triangular lattice.

For DHL, the wrapping probability $W(p, L)$ is plotted against $p$ for different values of $L$  in Fig. \ref{wp}(b). Unlike DTL, the curves do not intersect at a critical threshold; instead, they all meet at the threshold $p_c$. Employing least square intersection, we find $p_c = 0.4999 \pm 0.0012$ (indicated by the dashed line and marked by the cross on the $p$-axis). Note that, within the error bars, the critical threshold of DHL coincides with that of nearest-neighbour site percolation on the triangular lattice, $p_c = 1/2$ \cite{sykes1964exact,malarz2020site}. Corresponding to this $p_c$, the critical bond density on DHL is $\rho_c = 1/2$, whereas the ordinary bond percolation threshold on the honeycomb lattice $p_c = 1-2\sin(\pi/18) = 0.65270364466$ \cite{sykes1964exact}, which is significantly higher than the critical bond density on the DHL. 

In DHL, the perimeter bond cluster spans the lattice only at $p=p_c$ with the maximum wrapping probability, $\widetilde{W}[0] \approx 0.69$. Whereas for DTL, the wrapping probability of the perimeter bond clusters at $p=p_c$ (of DTL) is $\widetilde{W}[0] \approx 0.84$, and it approaches $1$ beyond $p=p_c$. The values of $\widetilde{W}[0]$ usually depend on the spanning conditions \cite{newman2000efficient}.  We have applied spanning conditions in vertical and horizontal directions on both square lattice and DTL to estimate $W(p, L)$ and found that $\widetilde{W}[0]\approx0.8$. However, the value of $\widetilde{W}[0]$ for DHL under the same spanning condition is found to be much less. Note that the perimeter bond cluster on the DL percolates at $p=p_c$ precisely when the unoccupied site clusters, with nearest-neighbour connections, on the PL depercolate.
%Thus, a special feature of MWP is that if the perimeter bond cluster percolates at $p_c$ on a dual lattice, then the nearest-neighbour site percolation occurs at $1-p_c$ on the corresponding primal lattice, independent of the lattice structure. 

In ordinary percolation, the wrapping probability is expected to be a Heaviside step function at $p=p_c$ in the limit $L\to\infty$. The wrapping probability of DTL in the limit $L\to\infty$ can be described by a product of two Heaviside step functions, as   
\begin{equation}
    \label{hf_DTL}
    W_{\rm{DTL}} (p) = H[p - p_c]H[(1-p_c)-p]
\end{equation}
where $H$ represents the Heaviside step function. Whereas, for DHL, the perimeter bond cluster percolates only at $p_c = 1/2$ with the maximum probability of $\widetilde{W}[0] \approx 0.69$. Hence, in the limit $L\to\infty$, the wrapping probability for DHL should be expressed as a product of two Heaviside step functions along with a coefficient $\widetilde{W}[0]$, and it is given by 
\begin{equation}
    \label{hf_DHL}
    W_{\rm{DHL}}(p) = \widetilde{W}[0]H[p - p_c]H[p_c - p].
\end{equation}
Since both the thresholds coincide at a single point $p=p_c$, $W_{\rm{DHL}}(p)$ can also be expressed as $W_{\rm{DHL}}(p) = \widetilde{W}[0]\delta(p-p_c)$, which is unusual for a percolation problem. Thus, the two wrapping probabilities of MWP on DTL and DHL exhibit different behaviour. In Fig. \ref{wp}(c), we plot the analytic forms of the wrapping probabilities,  $W_{\rm{DTL}}(p)$ and $W_{\rm{DHL}}(p)$, in the limit $L\to\infty$  given in Eqs. \ref{hf_DTL}, \ref{hf_DHL} as a function of $p$ in the range of $0 \leq p \leq 1$. It can be seen that in the case of DTL, there exist two critical thresholds $p_c = 0.30295$ and $1-p_c = 0.69705$. For $p_c<p<1-p_c$, the perimeter bond cluster percolates and the system exhibits a spanning phase between the two thresholds. Whereas, for DHL, the two thresholds $p_c$ and $1-p_c$ merge at $0.5$, and the system spans only at $p = 0.5$. There is no finite interval of $p$ over which the system remains in a spanning phase. Since spanning occurs only at this single critical point, we refer to it as the spanning point.

%----------------------------------------------------------
\begin{figure*}[ht]
\centerline{\hfill\includegraphics[width=0.31\linewidth]{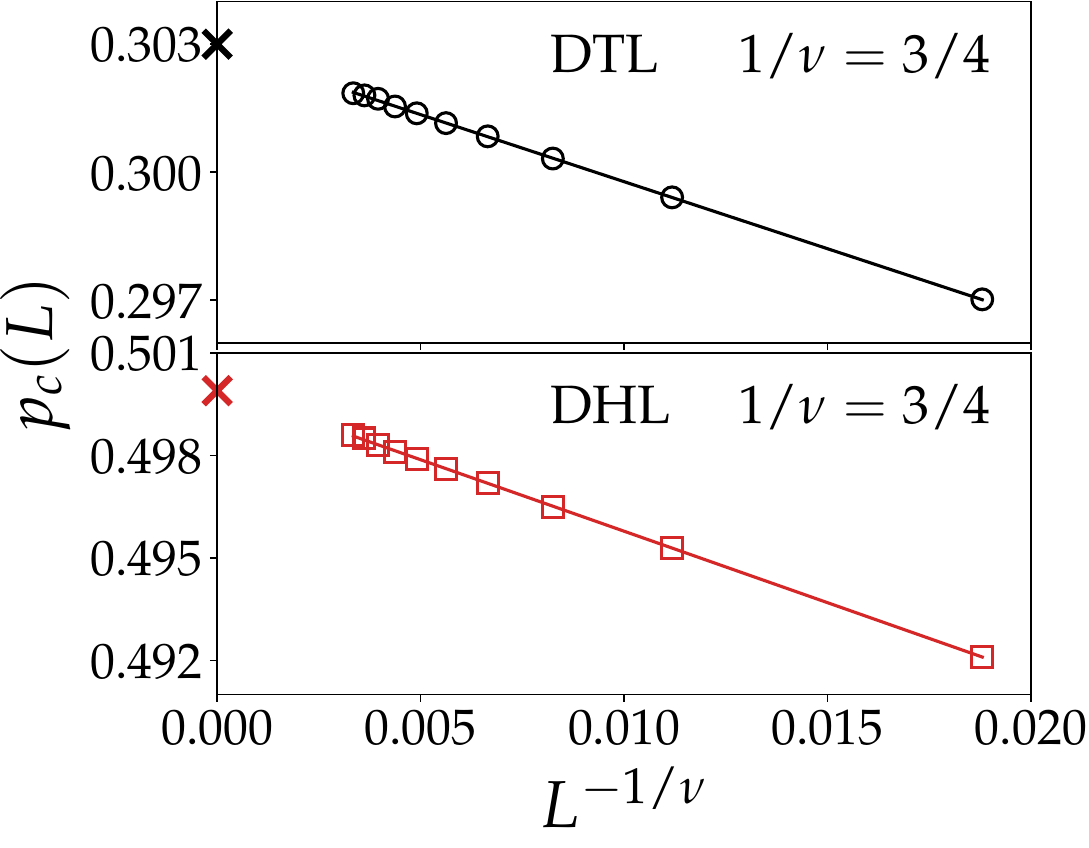}\hfill\hfill                     \includegraphics[width=0.33\linewidth]{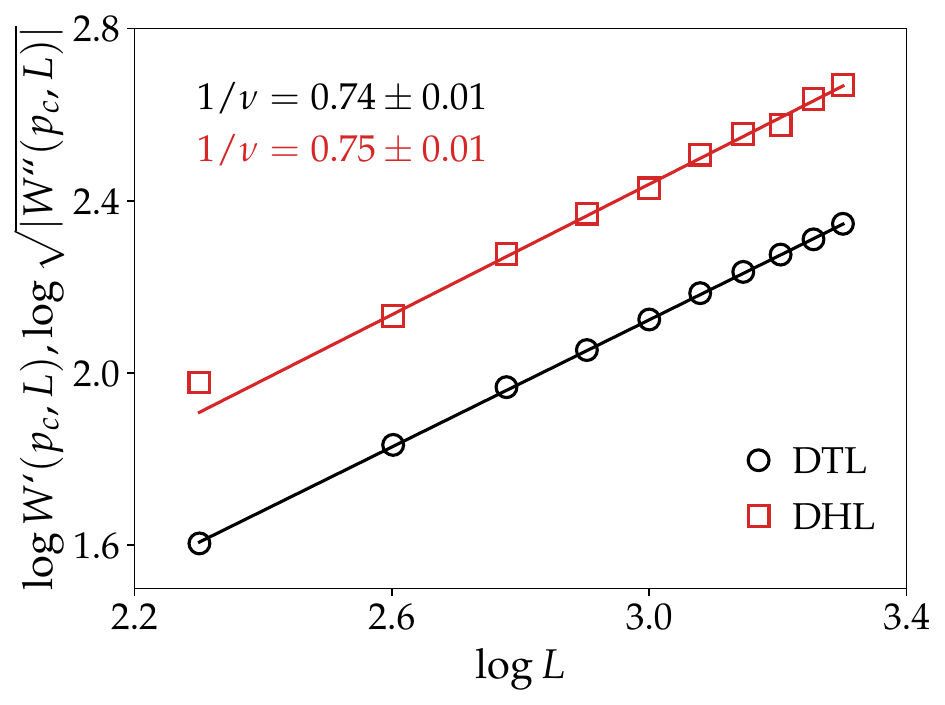}\hfill
        \includegraphics[width=0.33\linewidth]{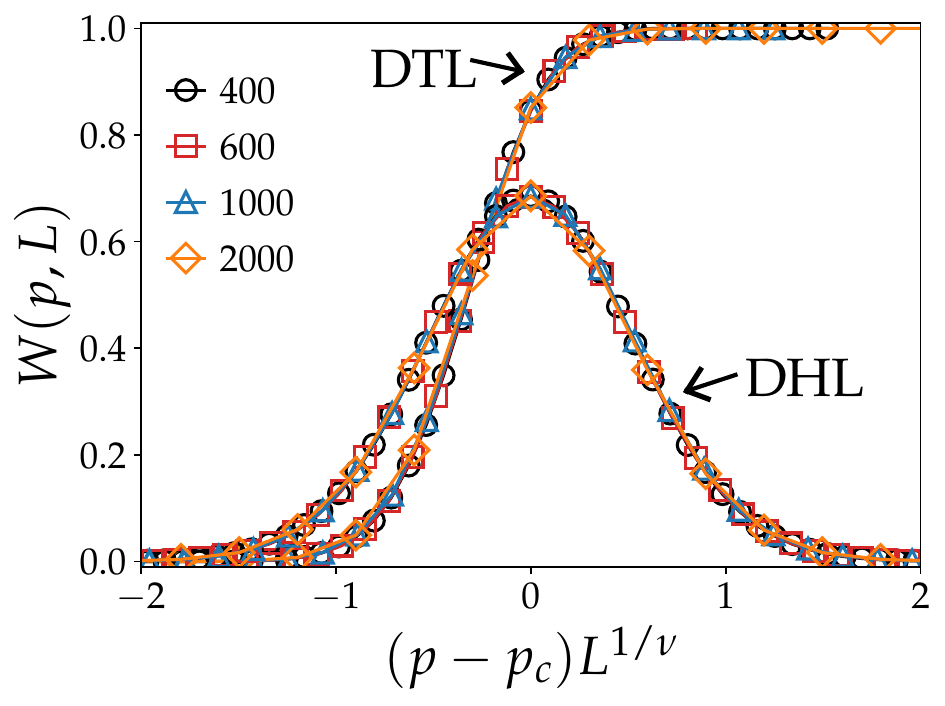}\hfill}
       \centerline{\hfill (a)\hfill\hfill (b)\hfill\hfill (c)\hfill} 
\caption{\label{fss_wp} In (a), we plot $p_c(L)$ against $L^{-1/\nu}$ with $\nu=4/3$. The upper and lower panels correspond to DTL and DHL, respectively. The critical threshold is found to be $p_c = 0.30294 \pm 0.00074$ for DTL (shown by the black cross) and $p_c = 0.49998 \pm 0.00069$ for DHL (shown by the red cross). In (b), we plot $W'(p_c, L)$ for DTL and $\sqrt{|W''(p_c,L)| }$ for DHL against $L$ in a double logarithmic scale. The prime ($'$) indicates the first derivative and ($''$) indicates the second derivative with respect to $p$. The solid line represents the regression line with slope $0.74\pm 0.01$ for DTL and $0.75\pm 0.01$ for DHL. In (c), we plot $W(p, L)$ against a scaled variable $x = (p-p_c)L^{1/\nu}$, with $1/\nu =3/4$ for both DTL and DHL.}
\end{figure*} 
%----------------------------------------------------------

%\subsection{Confirmation of $p_c$ and Estimation of $\nu$}
\subsection{ Determination of \texorpdfstring{$\nu$}{nu} and scaling form of \texorpdfstring{$\widetilde{W}[x]$}{widetilde{W}[x]}}
\label{nu}
Although we have identified the critical thresholds, we would like to verify them using a finite-size analysis of the wrapping probability. The critical thresholds $p_c(L)$ for systems of finite sizes $L$ are identified by the condition that at $p_c(L)$ the wrapping probability is $W(p_c(L), L) = 1/2$. Then the Eq. \ref{eq-3-1}, becomes
\begin{equation}
    \widetilde{W}[(p_c(L)-p_c)L^{1/\nu}]=W(p_c(L),L)=\frac{1}{2}.
    \label{eq-3-2a}
\end{equation} 
Inverting Eq. \ref{eq-3-2a}, one obtains
\begin{equation}
p_c(L) = p_c + CL^{-1/\nu} 
\label{eq-3-3}
\end{equation}
where $C = \widetilde{W}^{-1}[1/2] $ is a constant and $\nu$ is the correlation length exponent. The value of $p_c(L)$ is obtained from the intersection of the line $W(p, L) = 1/2$ with the interpolated cubic spline curve of \(W(p, L)\). The value of $p$ corresponding to the intersection point is the value of $p_c(L)$. In Fig. \ref{fss_wp}(a), we plot $p_c(L)$ against $L^{-1/\nu}$ for DTL and DHL, respectively, assuming $\nu = 4/3$, the correlation length exponent of ordinary percolation. The least-squares fitting was used to fit a line through the data points. The intercept of the fitted line with the $y$-axis indicates the critical threshold $p_c$ (for $L\to\infty$). The critical threshold for DTL is found to be $p_c = 0.30294 \pm 0.00074$, and for DHL, it is $p_c = 0.49998 \pm 0.00069$. The errors include spline propagation errors and fitting errors. That not only confirms the values of $p_c$s but also shows a slight improvement in $p_c$ for DHL. Moreover, to find $p_c$s, we have used the value of $\nu$ as that of the ordinary percolation, which provides a very good linear fit to the data points. It suggests that the correlation length exponent $\nu$ for  MWP in both DTL and DHL (also on the square lattice) is the same as that of ordinary percolation. %{\color{red} For comparison, the exact critical threshold for ordinary bond percolation are $p_c = 2 \sin(\pi/18)$ on the triangular lattice and  $p_c = 1-\sin 2 (\pi/18)$ on the honeycomb lattice \cite{sykes1964exact}, while the corresponding exact thresholds for ordinary site percolation are $p_c = 1/2$ on the triangular lattice and $p_c = 1- \sin(\pi/3)/2$ on the honeycomb lattice \cite{stauffer2018introduction,christensen2005complexity}.  }

The value of the exponent $\nu$ can be evaluated by estimating the first and second derivatives of $W(p,L)$ at $p=p_c$. Following the Eq. \ref{eq-3-1}, the derivatives $W'(p_c,L)$ and $W''(p_c,L)$ at $p=p_c$ are given by
\begin{align}
    \label{eq-3-4}
    W'(p_c,L)=C_1 L^{1/\nu};\quad W''(p_c,L) = C_2 L^{2/\nu}
\end{align}
where ($'$) and ($''$) indicate the first and second derivatives with respect to $p$ and \(C_1=\widetilde{W}'[0]\) and \(C_2=\widetilde{W}''[0]\). For DTL, a third-order polynomial was fitted to $W(p)$ with $\Delta p = 0.0001$ around $p = p_c$. The derivative of the polynomial with respect to $p$ was then estimated at $p=p_c$ for various system sizes $L$. For DHL, the first derivative vanishes at $p = p_c$, since the function $W(p, L)$ attains a maximum at $p_c$. Therefore, the second derivative $W''(p_c, L)$ was estimated using a four-point central difference method for various system sizes $L$. In Fig. \ref{fss_wp}(b), we plot the $W'(p_c, L)$ for DTL and $\sqrt{|W''(p_c,L)|}$ for DHL against $L$ on a double logarithmic scale. From the slopes of the fitted lines, the values of $1/\nu$ are found to be $0.74 \pm 0.01$ for DTL and $0.75 \pm 0.01 $ for DHL, which are the same as that of the ordinary percolation. 

Finally, we verify the finite-size scaling form of $W(p, L)$ and confirm the values of $p_c$ and $1/\nu$ obtained. We use the thresholds for DTL and DHL to be $0.3029$ and $0.4999$, respectively, and set $1/\nu = 3/4$ for both. Following Eq. \ref{eq-3-1}, we plot $W(p, L)$ against a scaled variable $x = (p - p_c)L^{1/\nu}$ in Fig. \ref{fss_wp}(c) for DTL and DHL. A good collapse of $W(p,L)$ for several $L$ is observed against the scaled variable $x$. This data collapse confirms not only the precision of the values of $\nu$ and $p_c$ but also the accuracy of the scaling form of $W(p, L)$. Although the data collapse of $W(p, L)$ in DTL is shown for the range $0\leq p \leq 0.5$ with $p_c = 0.30295$, a similar collapse also occurs for the range $0.5\geq p \geq 1$ with $1-p_c = 0.69705$. In contrast, for DHL, both thresholds are at $p=p_c=1/2$, so the data collapse of $W(p, L)$ occurs on both sides of $p=1/2$. 

%----------------------------------------------------------
\begin{figure*}[ht]
\centerline{\includegraphics[width=0.48\linewidth,clip]{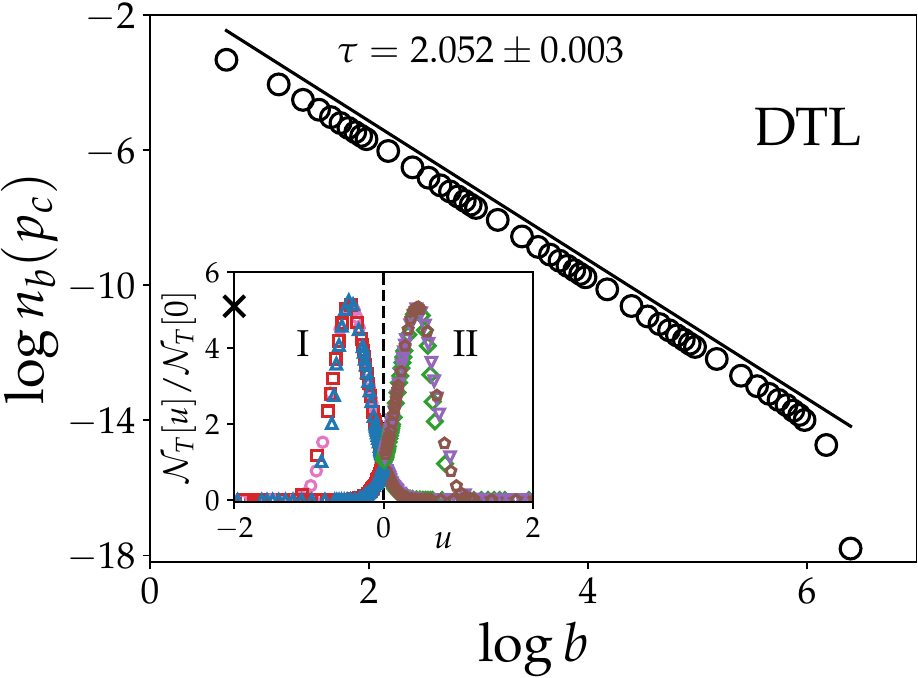}
            \includegraphics[width=0.48\linewidth]{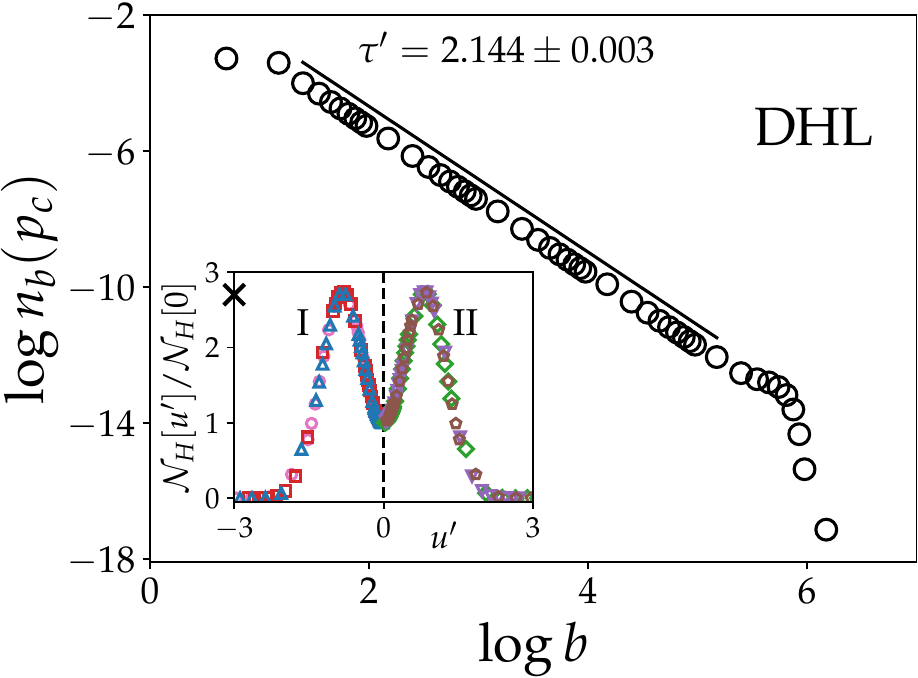}}
            \centering
            (a)\hspace{0.48\linewidth}(b)\\          
\caption{\label{clus_size} Plots of cluster size distribution of perimeter bond clusters $n_b(p_c)$ against the size of the perimeter bond cluster $b$ for (a) DTL and (b) DHL of lattice size $L = 2048$ are shown by the black circles on a double logarithmic scale. The black lines are guides to the eye with slopes $\tau = -187/91$ for DTL and $\tau' = -15/7$ for DHL. The insets show the collapse of $n_b(p)/n_b(p_c)$ against the scaled variables (a) $u=(p-p_c)b^\sigma$ for DTL and (b) $u'=(p-p_c)b^{\sigma'}$ for DHL near their corresponding critical thresholds. The symbols correspond to $p = 0.290,0.710$ (\LARGE{\color{Lavender}$\boldsymbol{\circ}$}\normalsize), $p = 0.295,0.705$ ({\color{red}$\boldsymbol{\square}$}), $p = 0.300,0.700$ ({\color{Cerulean}$\triangle$}), $p = 0.310,0.690$ (\LARGE{\color{Green}$\diamond$}\normalsize), $p = 0.315,0.705$ ({\color{RoyalPurple}$\boldsymbol{\triangledown}$}) and  $p = 0.320,0.710$ ({\color{Brown}$\boldsymbol{\pentagon}$}) for DTL. The symbols correspond to $p = 0.47$ (\LARGE{\color{Lavender}$\boldsymbol{\circ}$}\normalsize), $p = 0.48$ ({\color{red}$\boldsymbol{\square}$}), $p = 0.49$ ({\color{Cerulean}$\triangle$}), $p = 0.51$ (\LARGE{\color{Green}$\diamond$}\normalsize), $p = 0.52$ ({\color{RoyalPurple}$\boldsymbol{\triangledown}$}) and $p = 0.53$ ({\color{Brown}$\boldsymbol{\pentagon}$}) for DHL. The values of $\sigma$ and $\sigma'$ used are $36/91$ for DTL and $3/7$ for DHL.}
\end{figure*} 
%---------------------------------------------------------- 

\subsection{Cluster Size Distribution}
\label{cluster_size_dis}
The size of a perimeter bond cluster is the number of bonds $b$ present in that cluster. The cluster number density $n_b(p)$ is defined as the number of perimeter bond clusters of size $b$ per link of a lattice. It is given by
\begin{equation}
\label{eq-cls-1}
n_b(p) = \frac{N_b(p)}{3L^2}
\end{equation}
where $N_b(p)$ is the number of perimeter bond clusters of size $b$ and $3L^2$ is the total number of links. The scaling forms of the cluster size distributions on DTL and DHL are given by
\begin{align}
\label{eq-cls-2a}
n_b(p) &= b^{-\tau}\mathcal{N}_T[(p-p_c)b^{\sigma}]\\ \label{eq-cls-2b} n_b(p) &= b^{-\tau '}\mathcal{N}_H[(p-p_c)b^{\sigma '}]
\end{align}
where $\mathcal{N}_T [u]$ and $\mathcal{N}_H [u']$ are the scaling functions for the DTL and DHL, respectively. The variables $u = (p-p_c)b^\sigma$ and $u' = (p-p_c)b^{\sigma'}$ are the corresponding scaled variables, $\tau, \sigma$ and $\tau', \sigma'$ are the associated critical exponents of $n_b(p)$ for DTL and DHL respectively. At $p = p_c$, \(n_b(p)\) is given by 
\begin{equation}
\label{eq-cls-3}
n_b(p_c) \sim b^{-\tau};{\ \ } n_b(p_c) \sim b^{-\tau'}
\end{equation}
as $\mathcal{N}_T [0]$ and $\mathcal{N}_H [0]$ are constants for DTL and DHL, respectively. Clusters of different sizes are obtained from $10^5$ independent realisations on a lattice of size $L=2048$ and are distributed into different bins. In Fig. \ref{clus_size}(a), the size distribution function of DTL $n_b(p_c)$ is plotted against the size of the clusters $b$ on a double logarithmic scale. The slope of the distribution is found to be $\tau = 2.052 \pm 0.003$ for DTL. The black line serves as a guide to the eye and has a slope of $187/91$, corresponding to the cluster exponent of ordinary percolation. A similar behaviour was observed in the case of the square lattice, where the value of $\tau$ also belongs to the ordinary percolation universality class \cite{DAS2025130957}. In Fig. \ref{clus_size}(b), $n_b(p_c)$ is plotted against the size of the clusters $b$ on a double logarithmic scale for DHL. The slope of the distribution is found to be $\tau ' = 2.144 \pm 0.003$ for DHL. The black line serves as a guide to the eye, with a slope of $ 15/7$, corresponding to that of the ordinary percolation for hulls. It is evident from Fig. \ref{morphology} that, in the case of the DHL, the perimeter bond clusters on the DL are the hulls of the site clusters on the PL. In the absence of knots, the external hulls cannot connect with the internal hulls on the DHL. However, that is not the case for DTL or square lattice \cite{DAS2025130957}. Note that the hulls are a spontaneous outcome of the cluster-generating mechanism in MWP, rather than being extracted from any site/bond clusters, as in ordinary percolation.  

%----------------------------------------------------------
\begin{figure*}[ht]
\centerline{\hfill\includegraphics[width=0.33\linewidth,clip]{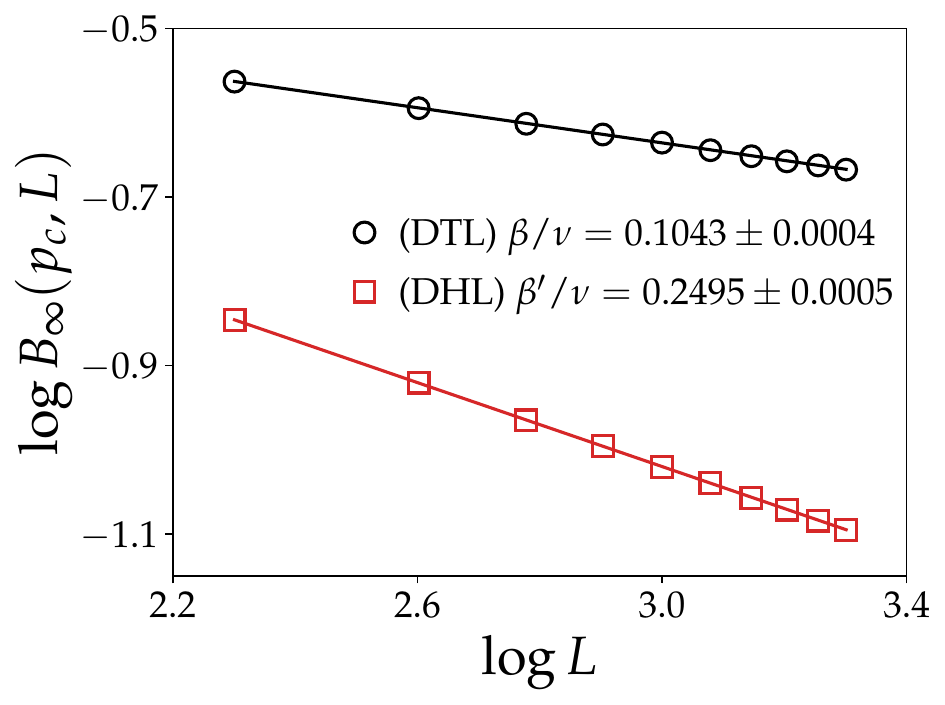}\hfill
            \includegraphics[width=0.33\linewidth,clip]{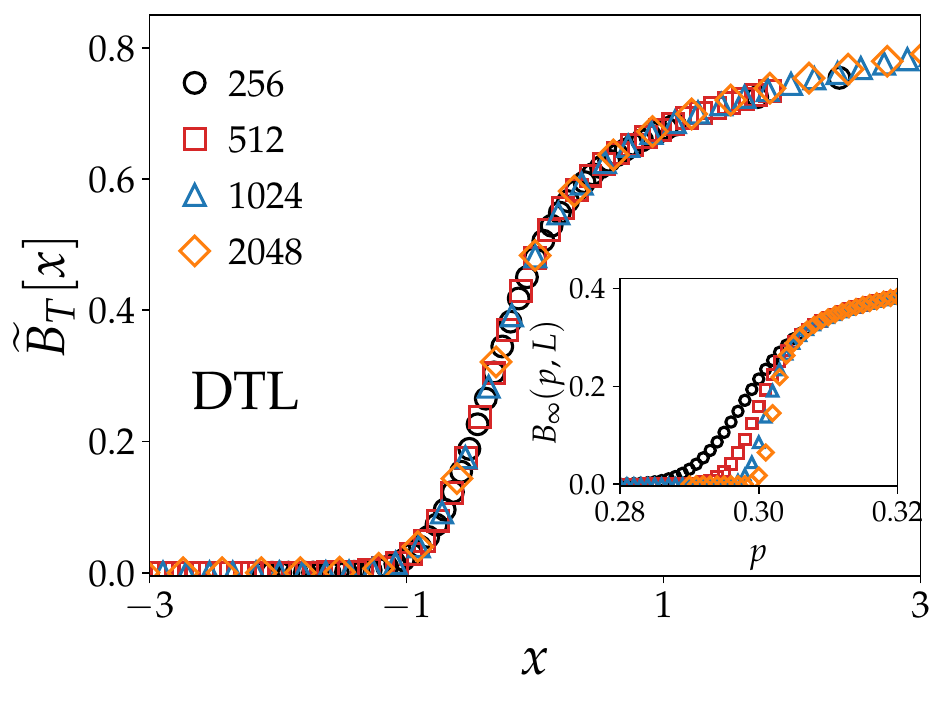}\hfill
            \includegraphics[width=0.33\linewidth,clip]{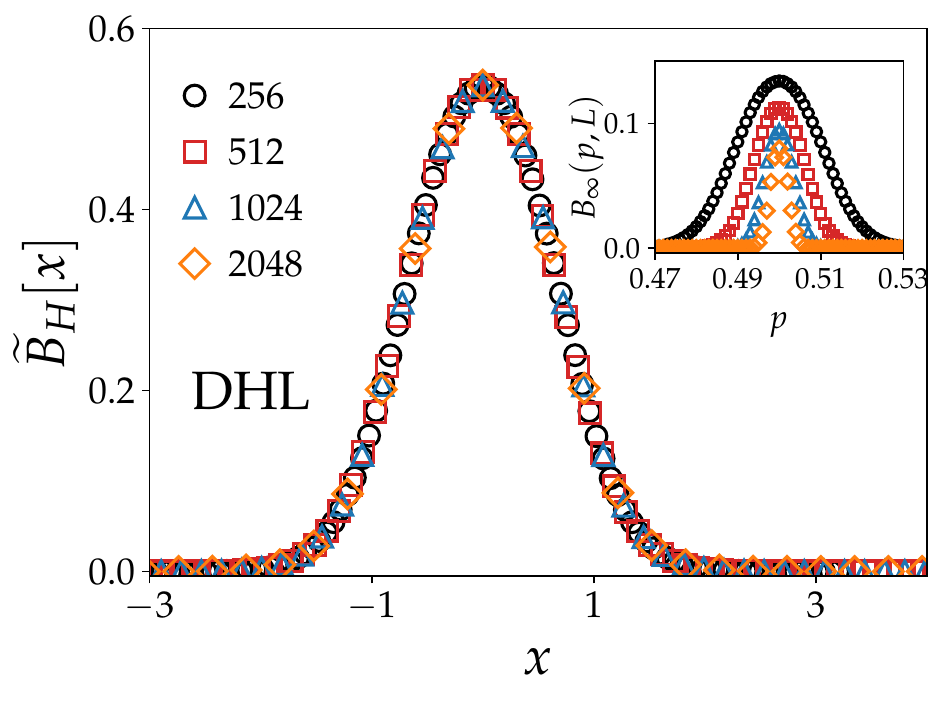}\hfill}
\centerline{\hfill(a)\hfill\hfill(b)\hfill\hfill(c)\hfill}
\caption{\label{fss_op}(a) Plot of the order parameter $B_\infty(p_c, L)$ at $ p = p_c$ as a function of lattice size $L$ on a double logarithmic scale for DTL and DHL. The slope indicates the ratio of critical exponents $\beta/\nu = 0.1043 \pm 0.0004$ for DTL and $\beta'/\nu = 0.2495\pm 0.0005$ for DHL. The scaled order parameter (b) $B_\infty(p,L)L^{\beta/\nu} = \widetilde{B}_T[x]$ for DTL and (c) $B_\infty(p,L)L^{\beta'/\nu}= \widetilde{B}_H[x]$ for DHL is plotted against the scaled variable $x = (p-p_c)L^{1/\nu}$ for various lattice sizes. The exponent values $1/\nu = 3/4$, $\beta/\nu = 5/48$ and $\beta'/\nu = 1/4$ are used. The inset shows $B_\infty(p, L)$ as a function of $p$ for (b) DTL and (c) DHL. }
\end{figure*} 
%----------------------------------------------------------

Once we know $\nu$ and $\tau$, all the other exponents can be found using the scaling relations, as they are not all independent. The value of $\sigma$ and $\sigma '$ is particularly of interest here, as it is required to verify the scaling functions $\mathcal{N}_T$ and $\mathcal{N}_H$, respectively. The relations between $\sigma,\sigma'$, $\tau, \tau'$ and $\nu$ can be established  as,
\begin{equation}
    \label{eq-cls-4}
    \sigma = \frac{\tau - 1}{d\nu}; \quad  \sigma' = \frac{\tau'  - 1}{d\nu}
\end{equation}
by utilising the hyper-scaling relations \(\tau-1=d/d_f\ \text{and}\ 1/\sigma=\nu d_f\) \cite{fisher1967theory, STAUFFER19791}. The value of $\sigma = 36 /91$ and $\sigma' = 3/7$ is obtained from Eq. \ref{eq-cls-4} using $\nu = 4/3$, $\tau = 187/91$ and $\tau' = 15/7$. It can be noticed that the values of $\sigma$ and $\sigma'$ correspond to the respective exponents for cluster ordinary percolation and hull ordinary percolation \cite{stauffer2018introduction}. In order to verify the form of the scaling functions $\mathcal{N}_T$ and $\mathcal{N}_H$, we look at the normalized scaling functions as,
\begin{equation}
    \label{eq-cls-5}
    \frac{\mathcal{N}_T[u]}{\mathcal{N}_T[0]}=\frac{n_b(p)}{n_b(p_c)};\quad \frac{\mathcal{N}_H[u']}{\mathcal{N}_H[0]}=\frac{n_b(p)}{n_b(p_c)}
\end{equation}
for both DTL and DHL, respectively. In the inset of Fig. \ref{clus_size}(a), we plot the ratio $\mathcal{N}_T[u]/ \mathcal{N}_T[0]$ for DTL against the scaled variable $u=(p-p_c)b^\sigma$ for several values of $p$ around $p_c=0.30295$ as well as around $1-p_c=0.69705$. The collapsed curve I is for the perimeter bond clusters that appear mostly around the occupied sites, and the collapsed curve II is for the perimeter bond clusters that appear mostly around the unoccupied sites. A collapse occurred in both datasets, providing verification of $\sigma = 36/91$. The two curves intersect at $(0,1)$ corresponding to $p=p_c$ (or $1-p_c$). The region $u>0$ of the curve I and the region $u<0$ of the curve II correspond to the spanning phases associated with critical thresholds $p_c$ and $1-p_c$. The maximum value of the curves is $\approx 5.1$, indicated by the cross, and is consistent with the cluster ordinary percolation case. In the inset of Fig. \ref{clus_size}(b), we plot the ratio $\mathcal{N}_H[u]/\mathcal{N}_H[0]$ for DHL against the scaled variable $u'=(p-p_c)b^{\sigma'}$ for several values of $p$ around $p_c=0.5$. The collapsed curve I corresponds to $p < p_c$, where the perimeter bond cluster appears mostly around the occupied sites, whereas the collapsed curve II corresponds to $p>p_c$, when the perimeter bond clusters appear mostly around unoccupied sites. A good collapse occurred for the dataset, which is a verification for $\sigma' = 3/7$. In contrast to the DTL case, the region $u'>0$ of the curve I and the region $u'<0$ of the curve II are absent, indicating the lack of a spanning phase in the DHL. The maximum height of the curves is $\approx 2.7$, as indicated by the cross. 
Note that the difference between the maximum value of the ratios $\mathcal{N}_T[u]/ \mathcal{N}_T[0]$
and $\mathcal{N}_H[u]/\mathcal{N}_H[0]$ indicates that $\mathcal{N}_T$ and $\mathcal{N}_H$ are different functions.

%----------------------------------------------------------
\begin{figure*}[ht]
\centerline{\hfill\includegraphics[width=0.33\linewidth,clip]{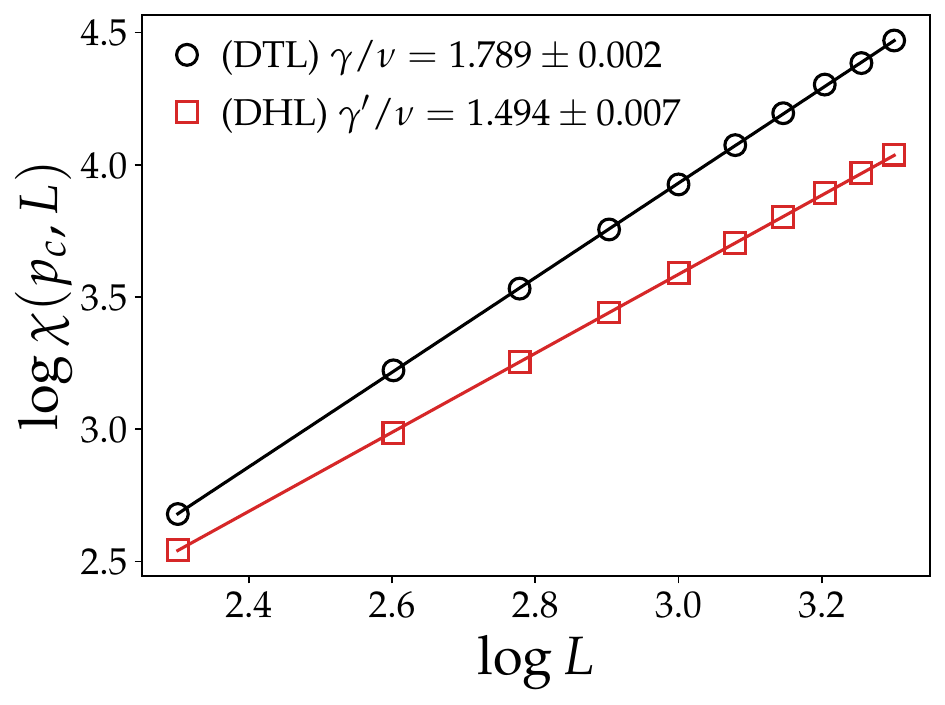}\hfill
            \includegraphics[width=0.33\linewidth,clip]{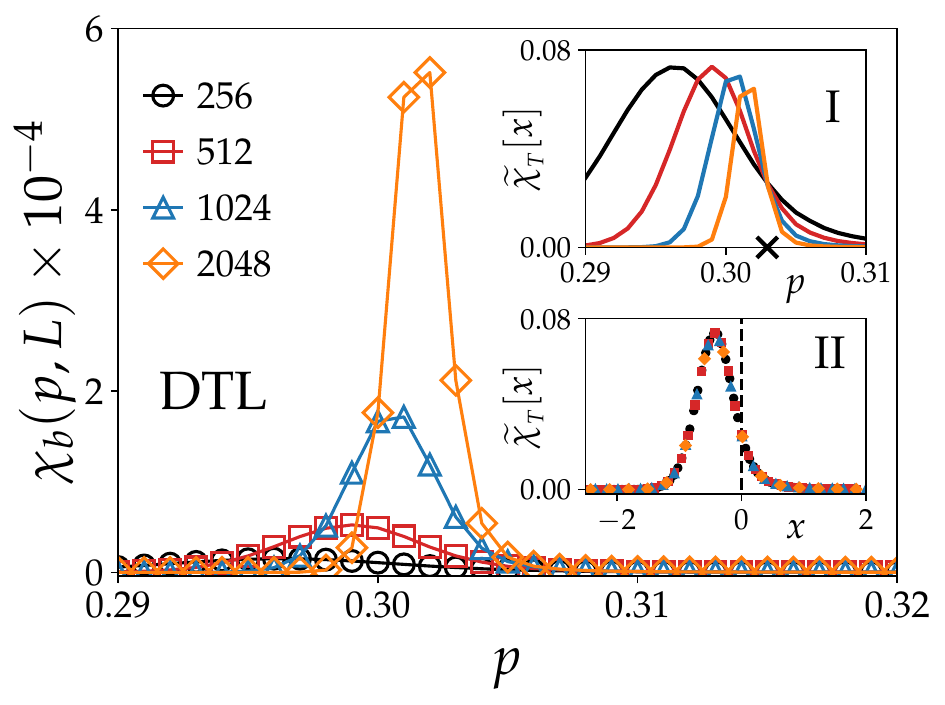}\hfill
            \includegraphics[width=0.33\linewidth,clip]{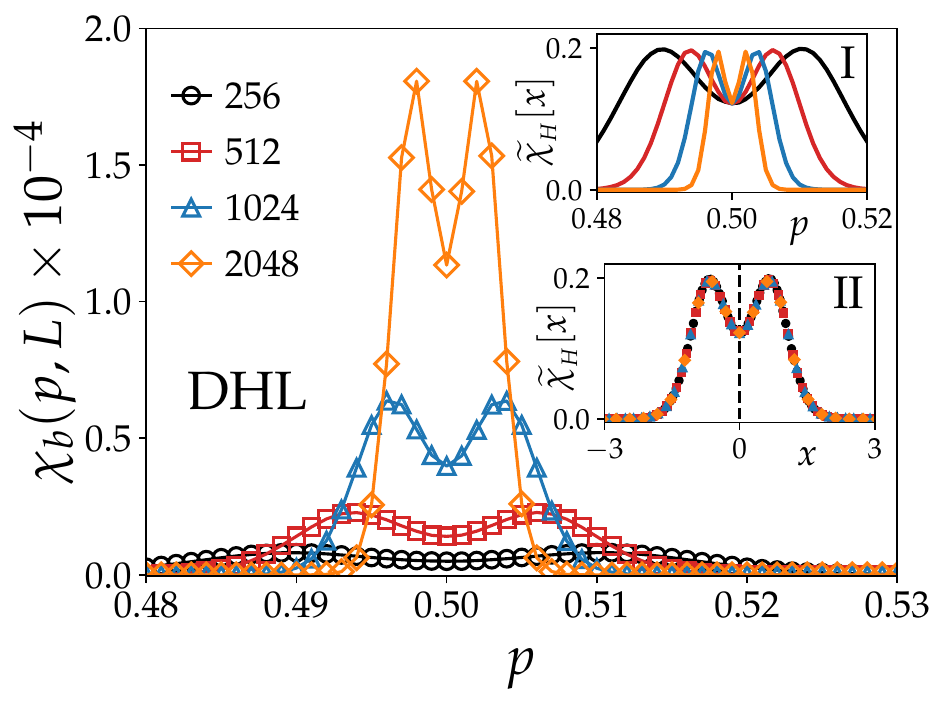}\hfill}
\centerline{\hfill(a)\hfill\hfill(b)\hfill\hfill(c)\hfill}
\caption{\label{fss_chi}(a) Plot of the fluctuation $\chi(p_c, L)$ at critical threshold $p_c$ as a function of lattice size $L$ on a double logarithmic scale for DTL and DHL. The slope indicates the ratio of critical exponents $\gamma/\nu = 1.789 \pm 0.002$ for DTL and $\gamma'/\nu = 1.494 \pm 0.007$ for DHL. The scaled fluctuation (b) $\chi(p,L)L^{\gamma/\nu}$ for DTL and (c) $\chi(p,L)L^{\gamma'/\nu}$ for DHL is plotted against the scaled variable $(p-p_c)L^{1/\nu}$ for various lattice sizes. The exponent values $1/\nu = 3/4$, $\gamma/\nu = 43/24$ and $\gamma'/\nu = 3/2$ are used. The inset shows $\chi(p, L)$ as a function of $p$ for (b) DTL and (c) DHL. }
\end{figure*} 
%----------------------------------------------------------

\subsection{Order parameter}
\label{FSS}

The order parameter ($B_\infty(p, L)$) of MWP is the probability of finding a bond on the largest (or spanning) perimeter bond cluster. It is defined as
\begin{equation}
    \label{eq-fss-1}
    B_\infty(p,L) = \rho(p) - \sum_b{\vphantom{\sum}}'bn_b(p),
\end{equation}
where $\rho(p)$ is the total bond density at a given $p$, and the second term represents the probability of finding a bond in finite clusters at that $p$ (the prime on the sum indicates exclusion of the largest cluster). The finite-size scaling forms of $B_\infty(p, L)$ for DTL and DHL are given by
\begin{align}
   B_{\infty}(p,L) &= L^{-\beta/\nu}\widetilde{B}_T\left[(p-p_c)L^{1/\nu}\right] \label{eq-fss-2} \\ B_{\infty}(p,L) &= L^{-\beta'/\nu}\widetilde{B}_H\left[(p-p_c)L^{1/\nu}\right] 
   \label{eq-fss-3}
\end{align}
where $\widetilde{B}_T[x]$ and $\widetilde{B}_H[x]$ are the scaling functions, $\beta$ and $\beta'$ are the critical exponents associated with $B_\infty(p)$ ($L\to\infty$) for DTL and DHL, respectively. Since at $p=p_c$, the scaling functions $\widetilde{B}_T[x = 0]$ and $\widetilde{B}_H[x = 0]$ reduce to constants, Eq. \ref{eq-fss-2} and \ref{eq-fss-3} can be represented as 
\begin{equation}
    \label{eq-fss-4}
   B_{\infty}(p_c,L) \sim L^{-\beta/\nu};{\ \ } B_{\infty}(p_c,L) \sim L^{-\beta'/\nu}
\end{equation}
for DTL and DHL, respectively. The ratio $\beta/\nu$ and $\beta'/\nu$ can be determined from Eq. \ref{eq-fss-4}. The values of $B_\infty(p_c, L)$ are estimated for several values of $L$ at $p_c=0.30295$ for DTL and at $p_c=0.5$ for DHL. In Fig. \ref{fss_op}(a), the values of $B_{\infty}(p_c, L)$ are plotted as a function of the lattice size $L$ on a double logarithmic scale for both DTL (shown by black circles) and DHL (shown by red squares). The solid lines are the regression lines. The slopes of the line yield the exponent ratios $\beta/\nu = 0.1043 \pm 0.0004$ and $\beta'/\nu = 0.2495 \pm 0.0005$ for DTL and DHL, respectively. The reported error in each exponent consists of the propagated statistical uncertainty and the least-squares fitting error. Notice that, the exponent ratio $\beta/\nu$ obtained for DTL is very close to the exact value $5/48$ for ordinary cluster percolation \cite{stauffer2018introduction,christensen2005complexity}, while that for DHL is close to the exact value $1/4$ for ordinary hull percolation \cite{ziff1986test,stauffer2018introduction}. The values of $B_\infty(p, L)$ are estimated for several values of $p$ in the vicinity of $p_c$ for different lattice sizes $L$ for both DTL and DHL. The inset of Fig. \ref{fss_op}(b) shows the variation of the $B_\infty(p, L)$ with $p$ for different lattice sizes $L$ for DTL. It can be seen that the transition becomes sharper with increasing $L$. The form of the scaling functions $\widetilde{B}_T[x]$ given in Eq. \ref{eq-fss-2} can be verified by utilising the ratio $\beta/\nu$ for DTL. In Fig. \ref{fss_op}(b), the scaling function $\widetilde{B}_T[x]=B_{\infty}(p, L)L^{\beta/\nu}$ (same as the scaled order parameter), is plotted as a function of the scaled variable $x = (p-p_c)L^{1/\nu}$ for different lattice sizes $L$. The data collapse well onto a single curve with $\beta/\nu=5/48$, the same as the ordinary cluster percolation, defining a unique function $\widetilde{B}_T[x]$. Note that the data for this collapse were collected for $p<0.5$, for which the perimeter bond clusters appear around the occupied sites. However, for $p>0.5$, most of the perimeter bond clusters appear around the unoccupied sites. A similar behaviour of $B_\infty(p, L)$ and the scaling function $\widetilde{B}_T[x]$ can be obtained at the other threshold $1-p_c = 0.69705$ for DTL. 

For DHL, the variation of $B_\infty(p, L)$ with $p$ for different lattice sizes $L$ is shown in the inset of Fig. \ref{fss_op}(c). It should be noted that at $p=0.5$, two thresholds, one for the perimeter bond clusters around the occupied sites for $p<p_c$ and the other for the perimeter bond clusters around the unoccupied sites for $p>p_c$, coexist. Thus, the $B_\infty(p, L)$ becomes sharper from both sides of the critical threshold, $p_c = 0.5$, as $L$ increases. The scaling function $\widetilde{B}_H[x] = B_\infty(p,L)L^{\beta'/\nu}$ is plotted as a function of the scaled variable $x = (p-p_c)L^{1/\nu}$ for different lattice sizes $L$ in Fig. \ref{fss_op}(c). A good collapse of data is observed in Fig. \ref{fss_op}(c), confirming the ratio $\beta'/\nu=1/4$ that of the ordinary hull percolation.

%----------------------------------------------------------
\begin{figure*}[ht]
  \centering{\includegraphics[width=0.32\linewidth]{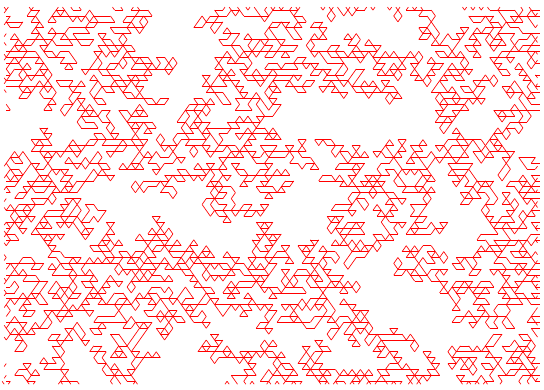} 
    \includegraphics[width=0.32\linewidth]{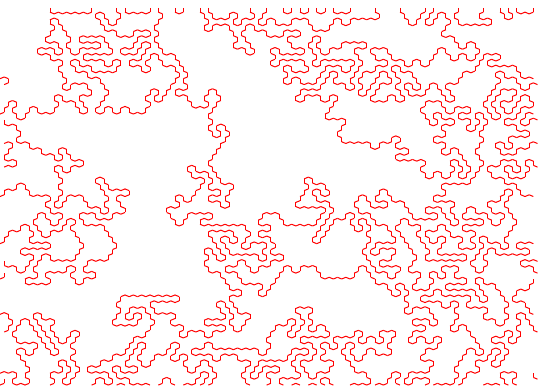} 
     \includegraphics[width=0.32\linewidth,clip]{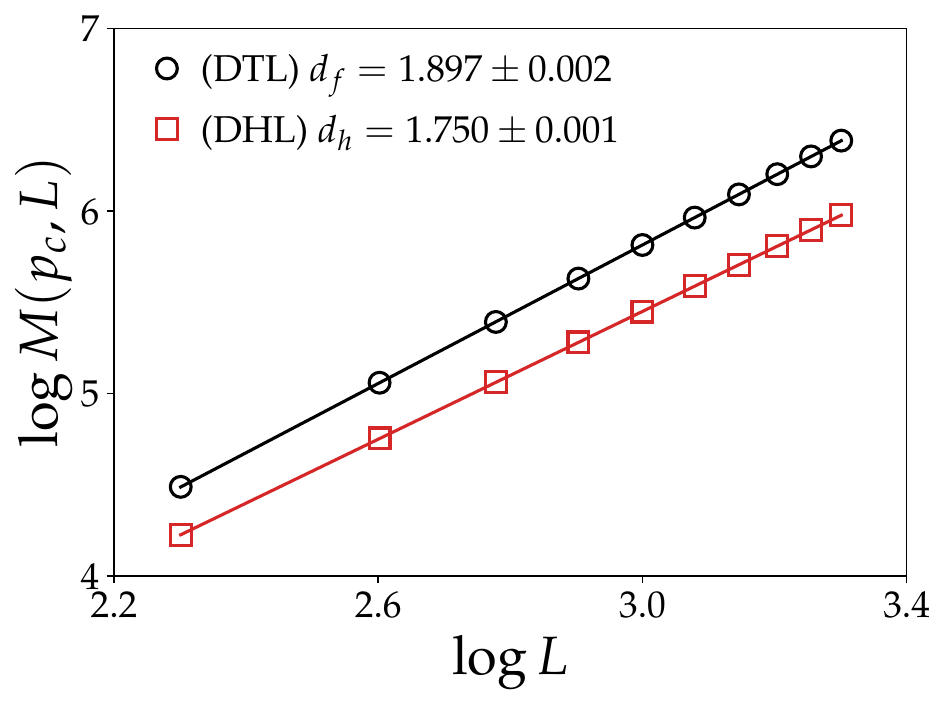}}\\ 
    \centerline{\hfill(a) DTL\hfill\hfill(b) DHL\hfill\hfill(c)\hfill} 
 \caption{\label{df} A spanning cluster for (a) DTL and (b) DHL at critical threshold $p = p_c$ on a lattice of size $L = 64$ is shown. (c) Plot of the mass $M(p_c, L)$ of the spanning perimeter bond cluster at critical threshold as a function of lattice size $L$ on a double logarithmic scale. The fractal dimension of the perimeter bond cluster is estimated as $d_f = 1.897 \pm 0.002$ for DTL (black circles) and $d_h = 1.750 \pm 0.001$ for DHL (red squares). The black and red solid lines denote the corresponding linear regression fits.}
\end{figure*}
%----------------------------------------------------------
\subsection{Fluctuation of order parameter}
\label{FOP}

Next, we consider the fluctuation in the order parameter of MWP. The fluctuation $\chi(p, L)$ in the order parameter $B_{\infty}(p,L)$ is defined as,
\begin{equation}
\label{eq-fss-5}
\chi(p,L) = L^{d}\left[\langle B_{\infty}^2(p,L)\rangle-\langle B_{\infty}(p,L)\rangle^2\right]
\end{equation}
where $d$ is the dimensionality of the system. The finite-size scaling form of $\chi(p, L)$ for DTL and DHL are given by,
\begin{align}
 \label{eq-fss-6} \chi(p,L) &= L^{\gamma/\nu}\widetilde{\chi}_T\left[(p-p_c)L^{1/\nu}\right] \\ \label{eq-fss-7} \chi(p,L) &= L^{\gamma'/\nu}\widetilde{\chi}_H\left[(p-p_c)L^{1/\nu}\right]  
\end{align}
where $\widetilde{\chi}_T[x]$ and  $\widetilde{\chi}_H[x]$ are the scaling function, $\gamma$ and $\gamma'$ are the critical exponents associated with $\chi(p)$ $(L \to \infty)$ for DTL and DHL, respectively. At $p = p_c$, the scaling functions $\widetilde{\chi}_T[0]$ and  $\widetilde{\chi}_H[0]$ reduce to constants. Consequently, the Eq. \ref{eq-fss-6} and (\ref{eq-fss-7}) can be expressed as,
\begin{equation}
    \label{eq-fss-8}
   \chi(p_c,L) \sim L^{\gamma/\nu};{\ \ } \chi(p_c,L) \sim L^{\gamma'/\nu}
\end{equation}
for DTL and DHL, respectively. The ratios of the exponents $\gamma/\nu $ and $\gamma'/\nu$ can be obtained from Eq. \ref{eq-fss-8} by evaluating $\chi(p_c,L)$ for different $L$ at $p_c=0.30295$ for DTL and at $p_c=0.5$ for DHL. In Fig. \ref{fss_chi}(a), the values of $\chi(p_c, L)$ are plotted against the lattice size $L$ on a double logarithmic scale for both DTL (shown by black circles) and DHL (shown by red squares). The solid lines are the regression lines. The slopes of the curves yield the ratios of the exponents $\gamma/\nu = 1.789 \pm 0.002$ and $\gamma'/\nu=1.494 \pm 0.007$. The error consists of the translated statistical error and the least-squares fitting error. For DTL, the value of $\gamma/\nu$ with the error bar is found to be $\approx 43/24$ that of the cluster ordinary percolation \cite{stauffer2018introduction}. Whereas, for DHL, the value of $\gamma'/\nu$ with the error bar is found to be $\approx 3/2$ that of the hulls of ordinary percolation clusters \cite{ziff1986test}. The scaling relations \(\gamma/\nu+2\beta/\nu=d, \text{and}\ \gamma'/\nu+2\beta'/\nu=d\) are verified within the error bars. 

The values of $\chi(p, L)$ are further estimated for several values of $p$ in the vicinity of $p_c$ for different lattice sizes $L$ for both DTL and DHL. The variation of $\chi(p, L)$ of DTL with $p$ is shown in Fig. \ref{fss_chi}(b) for several lattice sizes $L$. It can be seen that $\chi(p, L)$ diverges in the vicinity of $p_c$ as $L$ increases. The form of the scaling function $\widetilde{\chi}_T$ for DTL, given in Eq. \ref{eq-fss-6}, can be verified using the exponent ratio $\gamma/\nu$. Since the measured $\gamma/\nu$ is close to the exact ratio $43/24$, we use the exact value of $\gamma/\nu$ to study data collapse. In the inset I of Fig. \ref{fss_chi}(b), the scaling function $\widetilde{\chi}_T = \chi(p, L)L^{-\gamma/\nu}$ for DTL is plotted as a function of $p$ for different lattice sizes $L$. It can be seen that the curves of different $L$ intersect at $p=p_c$, indicating that the value of $\gamma/\nu$ is the same as that of the ordinary percolation cluster. The region $p > p_c$ of the curve in inset I corresponds to the spanning phase associated with $p_c$ (indicated by the cross). The scaling functions $\widetilde{\chi}_T$ for different $L$ are then plotted against the scaled variable $x=(p-p_c)L^{1/\nu}$ in the inset II of Fig. \ref{fss_chi}(b). A good data collapse onto a single curve is obtained with $\gamma/\nu = 43/24$ and $1/\nu = 3/4$, corresponding to the exponents of the cluster ordinary percolation. Note that the data used in the Fig. \ref{fss_chi}(b) is for $p < 0.5$, for which the perimeter bond clusters mostly surround the occupied sites. However, for $p > 0.5$, the perimeter bond clusters mostly surround the unoccupied sites. A similar behaviour of $\chi(p, L)$ and $\widetilde{\chi}_T[x]$ also appear around $1-p_c = 0.69705$.

For DHL, the variation of $\chi(p, L)$ with $p$ for different lattice sizes $L$ is shown in Fig. \ref{fss_chi}(c). With increasing $L$, $\chi(p, L)$ diverges as $p_c$ is approached from either side. Since at $p=0.5$, two thresholds coexist, $\chi(p<0.5, L)$ and $\chi(p>0.5, L)$ meet and terminate at $p=0.5$ as they can not have any spanning phase. The form of the scaling function $\widetilde{\chi}_H[x]$ for the DHL, given in Eq. \ref{eq-fss-7}, can be verified using the exponent ratio $\gamma'/\nu$. As the measured value of $\gamma'/\nu$ within the error bar is close to the exact value $3/2$ for hull percolation, the exact value is used for the data-collapse analysis. In the inset I of Fig. \ref{fss_chi}(c), the scaling function $\widetilde{\chi}_H[x] = \chi(p,L)L^{-\gamma'/\nu}$ for DHL is plotted as a function of $p$ for different lattice size $L$. All the curves meet one another at $p_c = 0.5$ from both sides and terminate. This indicates that the value of $\gamma'/\nu$ is consistent with that of the ordinary percolation hulls. Thus, MWP on DHL belongs to the universality of the ordinary percolation hull. Finally, the scaling function $\widetilde{\chi}_H$ for various $L$ are plotted against the scaled variable $x = (p - p_c)L^{1/\nu}$ in inset II of Fig. \ref{fss_chi}(c). A good data collapse onto a single curve is obtained by using $\gamma'/\nu = 3/2$ and $1/\nu = 3/4$, which are consistent with those of the ordinary percolation hull.  

\subsection{Fractal Dimensions}
\label{fractal}
Finally, we examine the fractal dimensions of the spanning clusters in these two cases. The spanning clusters generated for DTL and DHL on lattices of size $L = 64$ at their respective $p_c$ values are shown in Fig. \ref{df}(a) and \ref{df}(b), respectively. They appear quite distinct from each other. In the case of DTL, external perimeter bond clusters are connected to the internal perimeter bond clusters via knots. In contrast, in the case of DHL, the perimeter bond clusters cannot connect to one another because there are no knots. The tortuosity of the spanning perimeter bond cluster on DTL resembles the spanning site cluster on the primal honeycomb lattice. Whereas the spanning perimeter bond cluster on DHL resembles the exterior hull of the site cluster on the primal triangular lattice. 

At $p = p_c$, the average mass of the spanning cluster $M(p_c,L)$ should scale as,
\begin{equation}
    \label{eq-df}
    M(p_c,L) \sim L^{d_f};{\ \ } M(p_c,L) \sim L^{d_h}
\end{equation}
where $d_f$ and $d_h$ are the fractal dimensions of the spanning clusters of DTL and DHL, respectively. For both systems, the average mass of the spanning clusters is estimated at their respective $p_c$ values for different values of $L$. In Fig. \ref{df}(c), the average mass of the spanning cluster $M(p_c, L)$ is plotted against the lattice size $L$ on a double logarithmic scale. For DTL, the data are represented by black circles; for DHL, by red squares. The solid lines are the regression lines. For DTL, the slope of the line yields the fractal dimension $d_f = 1.897 \pm 0.002$. Whereas for DHL, the slope of the line yields the fractal dimension $d_h = 1.750 \pm 0.001$. The fractal dimension $d_f$ of the perimeter bond clusters in DTL is found close to $91/48$, which is the exact value of the fractal dimension of ordinary percolation cluster \cite{stauffer2018introduction,feder2013fractals}. In contrast, the fractal dimension $d_h$ of the perimeter bond clusters in DHL is found close to $7/4$, which is the exact value of ordinary percolation hull dimension \cite{voss1984fractal,weinrib1985new,Saleur1987}. 

Now, the values of the critical exponents and the fractal dimension of MWP in DTL (as well as in the square lattice \cite{DAS2025130957}) are the same as those of the ordinary cluster percolation. Whereas the critical exponents and the fractal dimension of MWP in DHL are the same as those of the ordinary hull percolation. It is worth noting that the construction of MWP is very different from that of ordinary bond percolation, even though they exhibit the same critical behaviour. The critical behaviour of MWP on both square \cite{DAS2025130957} and triangular lattices with coordination numbers $z=4$ and $z=6$, respectively, belongs to the ordinary cluster percolation universality class, whereas the same model on the honeycomb lattice with coordination number $z=3$ belongs to the ordinary hull percolation universality class. Therefore, MWP exhibits a crossover from hull universality class at $z=3$ to cluster universality class at $z \ge 4$. Generally, the universality class of percolation is independent of the lattice structure on a given space dimension $d$. However, MWP exhibits such a breakdown of universality. Although in the study of directed spiral percolation in two dimensions, a breakdown of universality was reported \cite{sinha2004breakdown} in a different context. A coordination number dependent critical behaviour was also suggested in a study of the zero-temperature random-field Ising model \cite{Thongjaomayum2013}.

\subsection{Site Cluster Envelopes}
\label{dhl_ext}

\begin{figure}[ht]
%  \centering{
%  \includegraphics[width=0.85\linewidth]{4.pdf}}\\
%\centering
%            (a)\\      
\centering{
  \includegraphics[width=0.98\linewidth]{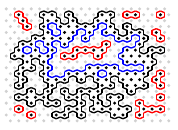}}\\ 
%\centering
%           (b)\\      
 \caption{\label{ext_clus} Illustration of cluster envelopes of a typical cluster morphology on DHL of size $L=20$ and $p = 0.5$. The black circles represent the occupied sites, and the grey circles represent the unoccupied sites on the primal triangular lattice. The black and blue lines represent the external and internal perimeter bond clusters, respectively, of the largest occupied site cluster; together, they form its cluster envelope. The red lines represent the perimeter bond cluster of the smaller occupied site clusters, with no holes inside, which correspond to their respective envelopes. }
 %The red arrows indicate visits to the nearest-neighbour unoccupied sites from an occupied site. The red arrows indicate visits to the nearest-neighbour unoccupied sites from the occupied sites. The numbers next to an unoccupied site indicate how many times that site is visited during the burning of the entire cluster of occupied sites.}}
\end{figure}

In the previous sections, it was observed that the perimeter bond clusters in DHL represent the hull of the occupied (or unoccupied) site clusters in the primal triangular lattice. However, to obtain the properties of the site clusters, one needs to capture the full geometry of each cluster. To obtain the geometry of a site cluster in DHL, one needs to find the envelope of the site cluster rather than the perimeter bond cluster in this case. A typical cluster morphology of the occupied sites (black circles) is generated with $p=0.5$ on the primal triangular lattice of size $L=20$ as shown in Fig. \ref{ext_clus}. The figure shows a large cluster of occupied sites (of size $s=132$) with many holes, along with $10$ small clusters (of size $s\le 8$) without holes. The perimeter bond clusters surrounding the occupied site clusters without holes are shown by red lines. For these site clusters, the external perimeter bond cluster surrounding the site cluster itself is the envelope of that cluster. For the site cluster with holes, the envelope consists of the external perimeter bond cluster shown by black lines and the internal perimeter bond clusters shown by blue lines in Fig. \ref{ext_clus}. Here, an internal perimeter bond cluster refers to the hull of a hole formed by unoccupied sites within an occupied site cluster. So far, in the previous sections, the external and internal perimeter bond clusters associated with a single site cluster were considered separately, and each was counted as an independent perimeter bond cluster, contrary to the present case, where all such perimeter bond clusters combined as a single cluster envelope. 
%To determine the cluster envelopes (on the corresponding dual lattice), we propose a new burning algorithm for the site clusters on the primal lattice.

%----------------------------------------------------------
%----------------------------------------------------------

%The new burning algorithm requires a list to store the indices of burned sites and a counter to record the number of bonds at the interface between occupied and unoccupied sites. Starting from a corner of the lattice, a search for an occupied site is carried out. Once an occupied site is found, it is burned, and its index is added to the list. Sitting on this burnt site, all its nearest neighbours are checked to see whether they are occupied or unoccupied. If an unoccupied nearest neighbour is found, the counter is incremented by one, since a bond exists between this conjugate pair of occupied and unoccupied sites. If an unburnt occupied nearest neighbour is found, it is burned, and its index is added to the list. (If the primal lattice were a honeycomb lattice (i.e. in the case of DTL), one needs to check up to next to next nearest neighbours for unburnt occupied sites.) Starting with the second entry in the list, the burning process is repeated; the list and the counter are updated. Each entry is then visited sequentially, and the burning process continues as long as the list is not exhausted. Note that an unoccupied site can be visited more than once from all its occupied nearest neighbours.

To determine the cluster envelopes (on the corresponding dual lattice), the burning algorithm is applied to the site clusters on the primal lattice. The burning process is similar to the bond burning process mentioned in Sec. \ref{Results}. Instead of burning the bonds, we burn the sites of the clusters, and count the number of bonds at the interface instead of the number of burnt sites. During the burning process, whenever we encounter an unoccupied nearest neighbour from a root burnt site we count it as an interface bond between occupied and unoccupied sites. An unoccupied site can be visited more than once from all its occupied nearest neighbours. At the end of the burning process, the counter value is the size $b$ of the cluster envelope, the total number of bonds present on the internal and external perimeters. Applying this burning process to the cluster configuration shown in Fig. \ref{ext_clus}, one would find the envelope size distribution $(b,N_b)$: $(6,4)$, $(10,2)$, $(14,2)$, $(22,1)$, $(32,1)$, and $(406,1)$, where $b$ is the size of an envelope and $N_b$ is the number of envelopes with size $b$. However, if the bond cluster burning algorithm (as described in Sec. \ref{Results}) is applied, sixteen perimeter bond clusters would be found for this configuration. It can be seen from Fig. \ref{ext_clus} that in this burning, the bonds on both the internal and external perimeters are counted, whereas the Ziff walk \cite{ziff1984generation} provides only the hull.

\begin{figure}[ht]
\centerline{\includegraphics[width=0.48\linewidth,clip]{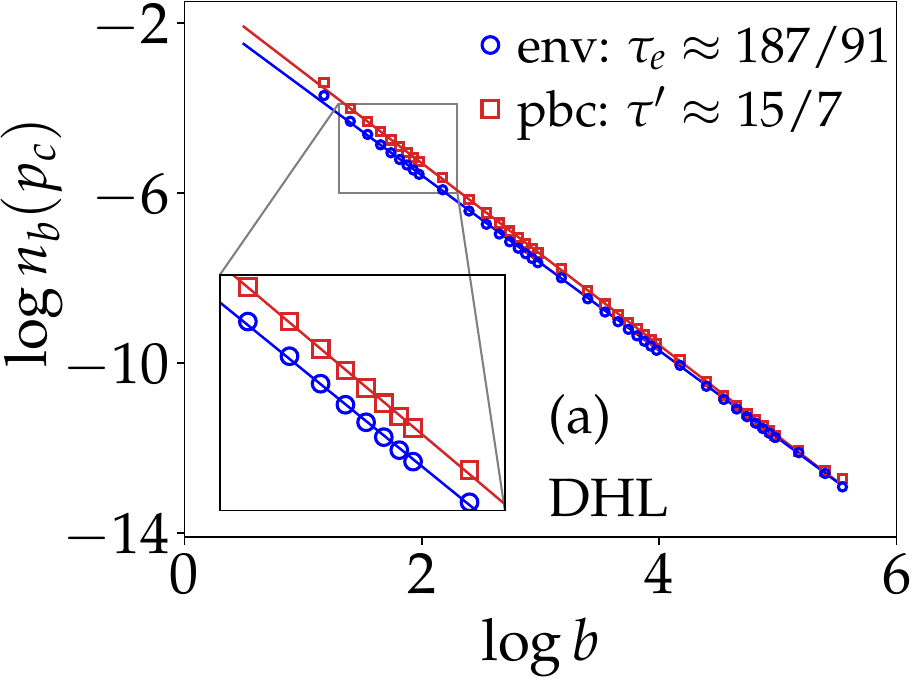}
\includegraphics[width=0.48\linewidth,clip]{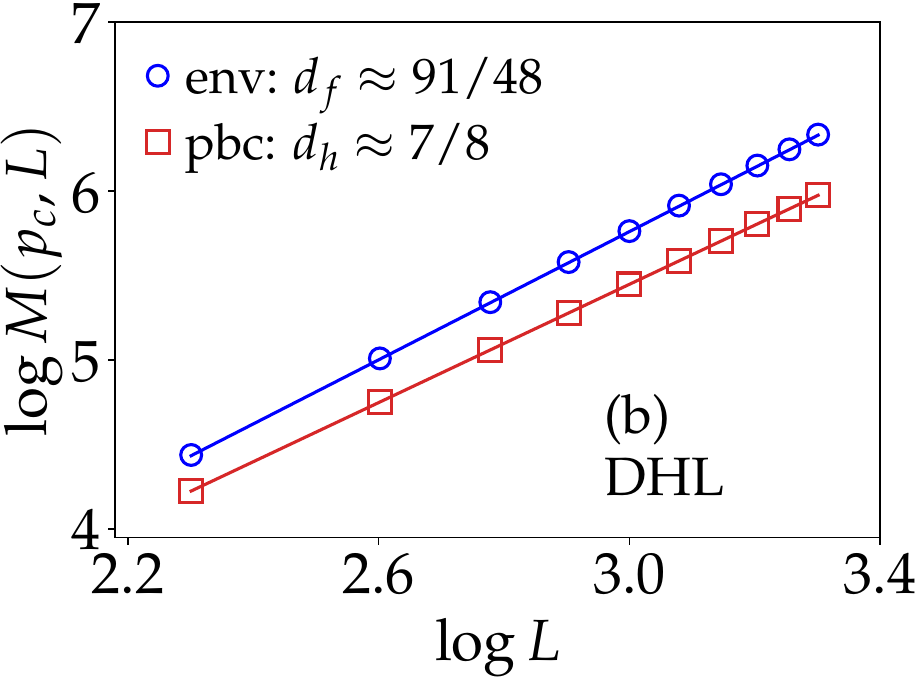}}
\centerline{\includegraphics[width=0.48\linewidth,clip]{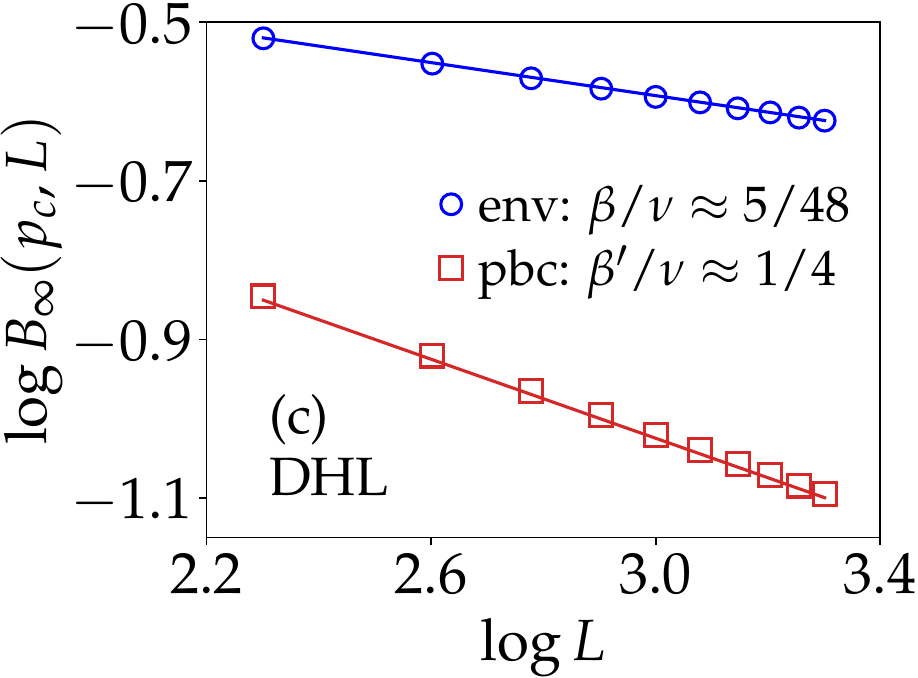}
\includegraphics[width=0.48\linewidth,clip]{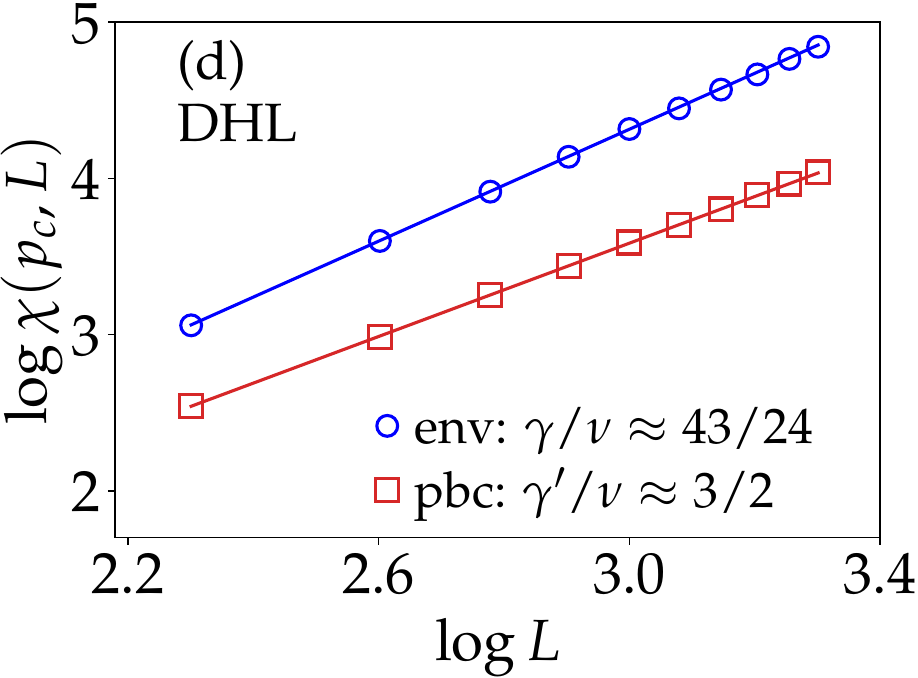} }
 
\caption{\label{cb_PTL} The data for cluster envelopes (env) are presented by blue circles, and that of the perimeter bond clusters (pbc) on DHL are presented by red squares. The solid lines in each plot, with slopes corresponding to the exact values of the critical exponents, are included as guides to the eye. (a) Plot of $n_b(p_c)$ as a function of envelope size $b$ on a lattice of size $L = 2048$. The inset highlights the difference in slopes. Plot of (b) $M(p_c, L)$, (c) $B_\infty(p_c, L)$, and (d) $\chi(p_c, L)$ of envelops and perimeter bond clusters against lattice size $L$. All plots are on double logarithmic scales.}
\end{figure}
%----------------------------------------------------------

Now, we study the critical properties of the envelopes and compare the results with the perimeter bond clusters on DHL. Since the cluster envelope and perimeter bond clusters span the lattice concurrently, the wrapping probability remains unchanged. Consequently, the critical threshold ($p_c = 0.5$) and the correlation length exponent ($1/\nu = 0.75$) are the same as those of the perimeter bond clusters in DHL. We have estimated the envelope size distribution $n_b(p_c)$, the fractal dimension $d_f$, the order parameter $B_\infty(p, L)$, and its fluctuations $\chi(p, L)$ to study the critical properties. 

The envelope size distribution $n_b(p_c)$ versus its size $b$ is plotted in Fig. \ref{cb_PTL}(a), shown by blue circles. The envelope size distribution exponent is found by linear regression to be $\tau_e = 2.060 \pm 0.005$, which is different from that of the perimeter bond clusters, $\tau' = 2.144 \pm 0.003$, shown by red squares. The solid blue and red lines are guides to the eye, with exact exponents for ordinary cluster percolation $187/91$ and hull percolation $15/7$. The exponent $\tau_e$ within the error bar is close to the exponent of the size distribution of the ordinary percolation cluster ($\tau = 187/91$). Since the critical exponents $\nu$ and $\tau$ of the cluster envelopes are the same as those of ordinary cluster percolation, the remaining critical exponents can be obtained from the scaling relations.

The mass $M(p_c, L)$ of the spanning envelope (in blue) is plotted against $L$ in Fig. \ref{cb_PTL}(b). The data for the spanning perimeter bond clusters are shown in red for comparison. It can be seen that $M(p_c, L)$ for the cluster envelopes scales differently from that for the perimeter bond clusters. The fractal dimension of the cluster envelopes is obtained by linear regression as $d_f =1.895 \pm 0.001$, which is consistent with that of ordinary percolation clusters. On the contrary, the fractal dimension of the perimeter bond cluster on DHL is that of ordinary percolation hulls ($d_h = 1.750 \pm 0.001$). The solid blue line with fractal dimension $91/48$ and the solid red line with hull dimension $7/4$ show that the estimated fractal/hull dimensions are close to the exact values. It can be seen that the mass of the envelopes is much larger than that of the perimeter bond cluster. It implicates the presence of many interior holes within the spanning clusters, which contribute substantially to the envelope mass. Knowing the fractal dimension $d_f$, one can determine the exponent ratios $\beta/\nu = d -d_f$ and $\gamma/\nu = 2d_f - d$ for the cluster envelopes. Nevertheless, for completeness, we determine these ratios of exponents numerically.

The order parameter $B_\infty(p, L)$ and its fluctuations $\chi(p, L)$ for the cluster envelopes are plotted against $L$ (in blue) in Fig. \ref{cb_PTL}(c) and \ref{cb_PTL}(d) respectively. The corresponding quantities for the perimeter bond clusters (in red) are also included for comparison. It can be seen that $B_{\infty}(p_c, L)$ and $\chi(p, L)$ for the cluster envelopes scale differently from those of the perimeter bond cluster. The exponent ratios $\beta/\nu = 0.1042 \pm 0.0003$ and $\gamma/\nu = 1.788 \pm 0.004$ for the envelopes are determined from the slopes of the linear regressions to the corresponding data of $B_\infty(p, L)$ and $\chi(p, L)$, respectively. The obtained value of the exponent ratio $\beta/\nu$ agrees, within the error bars, with the exact value $5/48$ for ordinary cluster percolation (shown by the blue solid line in Fig. \ref{cb_PTL}(c)) rather than with the corresponding value $1/4$ for ordinary hull percolation (shown by the red solid line). The estimated value $\gamma/\nu$ is consistent with the exact value of exponent ratios $43/24$ (blue solid line) for ordinary cluster percolation, but differs from the exact exponent ratio $3/2$ (red solid line) of ordinary hull percolation.

%Now we study the order parameter $B_\infty(p, L)$ and its fluctuations $\chi(p, L)$ for the cluster envelope. In the figure \ref{cb_PTL}(c), we plot the order parameter $B_\infty(p_c, L)$ for the cluster envelope and the perimeter bond cluster as functions of the lattice size $L$ on a double logarithmic scale. The different slopes of the two curves indicate that the envelope and perimeter bond cluster exhibit different scaling behaviour. From the slope of the fitted line through the data points of the envelope, we obtain the exponent ratio $\beta/\nu = 0.1042 \pm 0.0003$. This value is in close agreement with the corresponding exponent ratio for ordinary cluster percolation $\beta/\nu = 5/48$. In Fig. \ref{cb_PTL}(d), we plot the fluctuation of the order parameter $\chi(p_c, L)$ as a function of the lattice size $L$ for the cluster envelope, along with the corresponding data for the perimeter bond cluster. One can see that the slopes of the data points for the two curves differ. The exponent ratio $\gamma/\nu = 1.788 \pm 0.004$ is obtained from the slope of the fitted line for the envelopes, which is the same as that of ordinary cluster percolation $\gamma/\nu = 43/24$.

%----------------------------------------------------------
\begin{figure}[ht]
\centerline{\hfill\includegraphics[width=0.48\linewidth,clip]{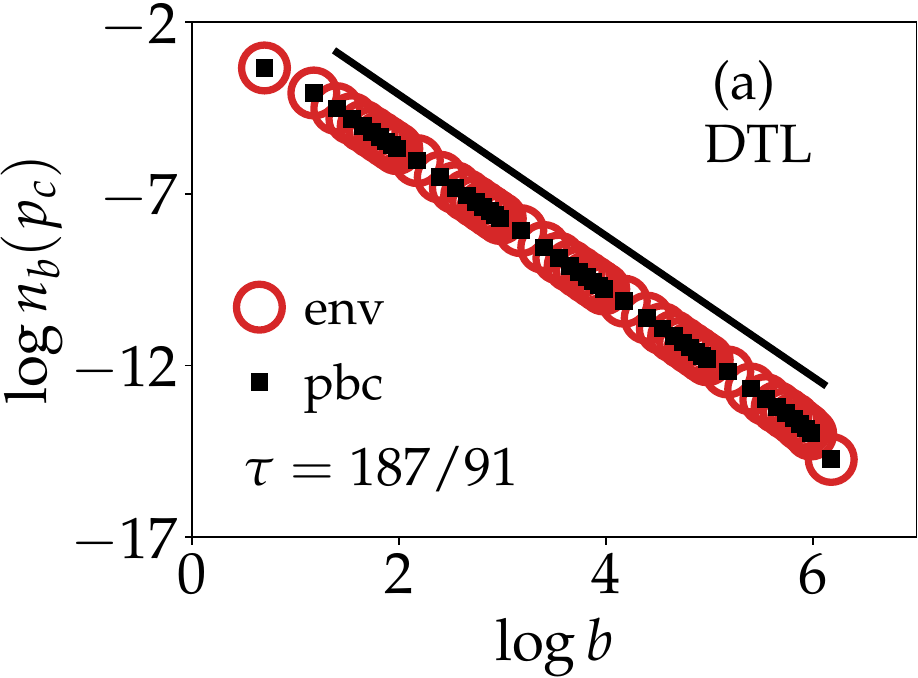}\hfill
\includegraphics[width=0.49\linewidth,clip]{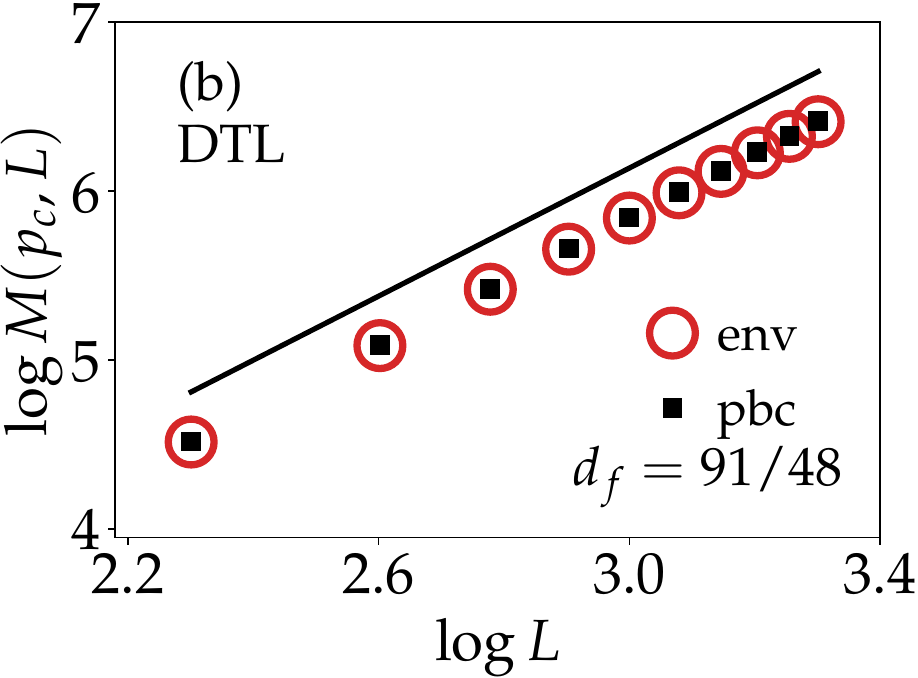}\hfill}
%          \centering 
\caption{\label{cb_PHL} The data for cluster envelopes (env) are presented by red circles, and that of the perimeter bond clusters (pbc) are presented by black squares on the DTL. Plot of (a) $n_b(p_c)$ versus envelope size $b$ and (b) $M(p_c, L)$ against lattice size $L$ on a double logarithmic scale. The solid lines in each plot are included as guides to the eye, with slopes corresponding to the exact values of the critical exponents.}
\end{figure}
%----------------------------------------------------------

However, on DTL, the perimeter bond clusters are found to capture the geometry of the underlying site cluster on the primal honeycomb lattice. Although it is possible that a small perimeter bond cluster can be trapped within a larger perimeter bond cluster, such cases are extremely rare. To demonstrate this, we estimate the envelope size distribution $n_b(p_c)$ as a function of envelope size $b$ and the mass $M(p_c, L)$ of the spanning envelopes as a function of lattice size $L$, and they are shown by red circles in Fig. \ref{cb_PHL}(a) and (b), respectively. For comparison, the corresponding perimeter bond cluster data is also shown by black squares in the respective plots. The two sets of data are indistinguishable within statistical uncertainty and exhibit the same critical behaviour as that of ordinary cluster percolation. The solid lines are guides to the eye, with slopes corresponding to the exact exponents, $\tau = 187/91$ in Fig. \ref{cb_PHL}(a) and $d_f = 91/48$ in Fig. \ref{cb_PHL}(b). This indicates that the contribution of the inner-perimeter bond clusters to the overall cluster statistics is negligible in DTL, whereas in DHL the internal perimeters make a significant contribution.

%In Fig.\ref{cb_PHL}, we plot the mass $M(p_c, L)$ of the spanning envelope (black circles) along with perimeter bond clusters (red squares) on DTL. It can be seen that both sets of data points are indistinguishable on the log scale. The black line with slope ($d_f = 91/48$) is a guide to the eye. The statistical properties of quantities such as the cluster size distribution, the order parameter and its fluctuations for the envelopes also exhibit the same critical behaviour as those of the perimeter bond cluster on DTL.

%In the DHL, the perimeter bond clusters appear on the honeycomb lattice, where they remain isolated and knotless. This is because the honeycomb lattice has a lower coordination number. The cluster envelopes, including the perimeter bond cluster on the outer front and the detached perimeter bond clusters on the internal front, together capture the geometry of the site cluster on the primal lattice. As a result, perimeter bond clusters exhibit hull-like properties, whereas cluster envelopes exhibit cluster-like properties.

\section{Conclusion}
\label{conclusion}

The scaling behaviour of perimeter bond clusters in the MWP exhibits a crossover from the percolation hull universality to the percolation cluster universality as the coordination number of the underlying dual lattice changes from \(z=3\ \text{to}\ z\ge 4.\) Such a coordination number dependent breakdown of universality is a rare phenomenon in the literature of percolation. The perimeter bond clusters represent the cluster of zero capillary force pore throats in the case of immiscible two-phase flow in a porous medium. Primarily, though the perimeter bond clusters obtained on different lattices are very different from those of ordinary bond percolation, they exhibit critical behaviour of percolation. Secondly, the perimeter bond clusters on DHL appear as hulls  which is a spontaneous outcome of the cluster-generating mechanism in MWP, rather than being extracted from any site/bond clusters, as in ordinary percolation. Finally, the critical behaviour of perimeter bond clusters on DTL ($z=6$) and DHL ($z=3$) belong to two different universality classes. On DTL, the higher coordination number allows many perimeter bond clusters to merge at a lattice site called a knot. As a result, the perimeter bond cluster captures the geometry of the site cluster on a primal honeycomb lattice. In contrast, the low coordination number of DHL prevents such merging of perimeter bond clusters, and each perimeter bond cluster remains a single tortuous line representing either an external or internal boundary of an occupied or unoccupied site cluster on the primal triangular lattice. Apart from their structural differences, the critical behaviour of perimeter bond clusters on DTL belongs to the cluster percolation universality, whereas that of the perimeter bond cluster on DHL belongs to the hull percolation universality. However, the cluster envelopes comprising both the external and internal perimeter bond clusters, exhibit the critical properties of ordinary cluster percolation on both DTL and DHL. It is interesting to note that on DHL the perimeter bond clusters and the envelopes (collection of external and all internal perimeters) yield two different universality classes of percolation in MWP. It indicates that the critical behaviour of percolation clusters is determined by the surface fluctuations rather than the whole mass of the clusters.

An interesting generalization of the MWP model is to consider the three-dimensional lattices. Do we find the same sub-universality classes in this case? The symmetries between primal and dual lattices we have seen in this work are specific to two dimensions, and the results in three dimensions will presumably be different.

\section*{Acknowledgment}
The authors are thankful to the Department of Science and Technology for the computational facilities under Grant No. SR/FST/P-11/020/2009 and Param Kamrupa of IIT Guwahati for the generous allowance of CPU hours.  This work was partly supported by the Research Council of Norway through its Centers of Excellence funding scheme, project number 262644, and the INTPART program, project number 309139. SS and AH furthermore acknowledge funding from the European Research Council (Grant Agreement 101141323 AGIPORE).

%-----------------------------------------------------------------
\bibliography{references}{}
%--------------------------------------------------------------------
\end{document}